\documentclass{jfm}

\usepackage{graphicx}
\usepackage{newtxtext}
\usepackage{newtxmath}
\usepackage{amssymb}
\usepackage[utf8]{inputenc}
\usepackage{natbib}
\usepackage{bm}
\usepackage{hyperref}
\hypersetup{
    colorlinks = true,
    urlcolor   = blue,
    citecolor  = black,
}
\graphicspath{{./}}

\newcommand{\bE}{\mathbf{E}}
\newcommand{\bU}{\mathbf{U}}
\newcommand{\bu}{\mathbf{u}}
\newcommand{\bn}{\mathbf{n}}

\newcommand{\be}{\mathbf{e}}

\newcommand{\bb}{\mathbf{b}}

\newcommand{\bhat}[1]{\hat{\mathbf{#1}}}
\newcommand{\eps}{\varepsilon}
\newcommand{\kap}{\kappa}
\newcommand{\dd}{\mathrm{d}}
\newcommand{\Order}[1]{O(#1)}
\newcommand{\DT}{\mathbf{D}^T}

\newcommand{\I}{\mathbf{I}}

\title{Shape-dependence of electrophoretic mobility: an AI-assisted perturbation analysis}

\author{Arkava Ganguly\aff{1}
 \and Ankur Gupta\aff{1}\corresp{\email{ankur.gupta@colorado.edu}}}

\affiliation{\aff{1}Department of Chemical and Biological Engineering, University of Colorado Boulder, Boulder, CO 80303, USA}

\begin{document}
\maketitle

\begin{abstract}

The electrophoretic mobility of a spherical particle is well understood, yet how particle shape modifies this mobility at arbitrary Debye length remains an open question. Here, we compute the electrophoretic mobility of a nearly spherical particle whose surface is modified through small axisymmetric spherical harmonic shape perturbations, at an arbitrary ratio of particle size to Debye length $\kappa a$. Using a volume-integral formulation combined with domain perturbation techniques, we derive a universal shape correction coefficient $\sigma_2(\kappa a)$ such that the mobility along the axis of symmetry takes the compact form $C_\parallel = f_H(\kappa a)\,[1 + \varepsilon\,c_2\,\sigma_2(\kappa a)]$, where $f_H$ is Henry's function. We show that $\sigma_2$ interpolates between $+1/5$ in the thick-double-layer (H\"{u}ckel) limit, governed solely by the Stokes drag correction, and zero in the thin-double-layer (Smoluchowski) limit. The perturbation theory agrees quantitatively with exact spheroidal solutions for both prolate and oblate orientations present in literature. A key finding is that only the $P_2$ (quadrupolar) component of the particle shape affects the mobility at leading order; higher harmonics are electrophoretically silent.  The physical formulation, the identification of the relevant asymptotic regime, and the interpretation of the results were carried out by the authors; the supporting perturbation algebra, numerical computations, and figures were developed with AI assistance (Claude, Anthropic) under close author verification. The AI made mistakes, such as fabricating a coefficient and forcing an interpolation, which we flagged and corrected. We discuss the role and limitations of AI in theoretical research, with representative prompts provided in an appendix.
\end{abstract}

\section{Introduction}
\label{sec:introduction}

Electrophoresis, the motion of a charged particle in a liquid electrolyte under an imposed electric field, is one of the oldest and most widely studied phenomena in colloidal science. Its origins trace back to the early nineteenth century, and its modern applications span the separation of biomacromolecules \citep{heller2001}, the measurement of zeta potentials \citep{doane2012}, microfluidic manipulation of particles and cells \citep{velegol2016}, and the characterization of surface charge in complex environments. The central quantity is the electrophoretic mobility, defined as the ratio of the particle velocity to the applied field strength. Over the last century, a large body of theoretical, computational, and experimental work has sought to understand how the mobility depends on the properties of the particle and the surrounding electrolyte. A comprehensive review of the historical development can be found in \citet{anderson1989}, and modern advances related to nonlinear electrophoresis are described in \citet{khair2022}.

Much of this theoretical progress has focused on the canonical geometry of a sphere. In the limit of a thin Debye layer ($\kap a \gg 1$, where $a$ is the particle radius and $\kap^{-1}$ is the Debye length), the Helmholtz--Smoluchowski formula gives the electrophoretic velocity as $\bU = (\eps_f \zeta / \mu)\bE_\infty$, where $\eps_f$ is the fluid permittivity, $\zeta$ is the zeta potential,  $\mu$ is the viscosity, and $\bE_\infty$ is applied electric field \citep{smoluchowski1921}. In the opposite limit of a thick Debye layer ($\kap a \ll 1$), \citet{huckel1924} showed that the mobility is reduced by a factor of $2/3$ relative to the Smoluchowski result, reflecting the balance between the electrostatic driving force and the Stokes drag. \citet{henry1931} bridged these two limits by computing the mobility at arbitrary $\kap a$ within the linearized Poisson--Boltzmann (Debye--H\"{u}ckel) approximation, yielding the well-known interpolation function $f_H(\kap a)$ that increases monotonically from $2/3$ to $1$. Extensions beyond the Debye--H\"{u}ckel regime, accounting for the distortion of the double layer during particle motion (the so-called relaxation effect), were developed by \citet{obrien1978} through a landmark numerical computation that revealed a mobility maximum at sufficiently large zeta potentials. More recent work has addressed the fully nonlinear regime at large applied fields, where the mobility itself becomes field-dependent \citep{khair2022}.

A remarkable (and a constraining feature for separation applications) of the Smoluchowski result is its independence of particle shape. \citet{morrison1970} proved that the electrophoretic velocity $\bU = (\eps_f \zeta / \mu)\bE_\infty$ holds for an insulating particle of \textit{any} shape, provided that the Debye layer is thin compared to all local radii of curvature and the zeta potential is uniform. The key insight is that the velocity field $\bu = (\eps_f \zeta / \mu)\bE$ is itself irrotational and satisfies both the Stokes equation and the slip boundary condition, regardless of the particle geometry. This result was independently confirmed by \citet{teubner1982}. As a consequence, shape effects on the electrophoretic mobility can only arise when the thin-Debye-layer assumption is relaxed.

The effect of particle shape at finite Debye length has received comparatively less attention. The most complete analytical treatment is the work of \citet{yoon1989}, who obtained exact solutions for spheroidal particles by solving the linearized Poisson--Boltzmann equation in spheroidal coordinates using spheroidal wavefunctions. Their results revealed several important features. First, the mobility of a spheroid depends on its orientation relative to the applied field. For a prolate particle, motion along the symmetry axis yields a different mobility than motion perpendicular to it. Second, for certain orientations (a prolate spheroid with the field perpendicular to its axis, or an oblate spheroid with the field along its axis), the mobility is not monotonic in $\kap a$ but exhibits a minimum at intermediate double-layer thicknesses, reflecting a subtle competition between electrostatic and hydrodynamic retardation forces. Third, when the mobility is averaged over all orientations (as would be measured for a randomly oriented ensemble), the result deviates only modestly from the sphere, and the orientation-averaged H\"{u}ckel-limit mobility is exactly $2/3$ for all spheroids, regardless of aspect ratio. Beyond spheroids, computational studies have addressed the electrokinetics of rod-like and slender particles \citep{solomentsev1994, ohshima1996}, but to the best of our knowledge, no general analytical framework for arbitrary shape effects on electrophoretic mobility exists.

The shape-independence of the Smoluchowski result implies that the question of how shape affects electrophoresis is inextricably tied to finite double-layer effects. This coupling presents a methodological challenge. Much of the modern literature on phoretic mobility calculations, particularly for self-propelled particles, relies on the slip velocity formulation popularized by \citet{stone1996}. In this framework, the thin-Debye-layer approximation is invoked to replace the electrokinetic body force with an effective slip velocity at the particle surface, after which the velocity is obtained from a surface integral using the Lorentz reciprocal theorem. This approach is powerful and widely used \citep{stone1996, masoud2019}, but it is restricted to thin Debye layers and therefore cannot capture shape-dependent corrections to the mobility. To study the interplay between shape and finite double-layer thickness, one must instead work with the full volume integral formulation, in which the body force throughout the fluid is retained explicitly.

In this work, we employ the unified mobility expression derived by \citet{ganguly2024}, building on the framework of \citet{brady2021}, to compute the electrophoretic mobility of a nearly spherical particle at arbitrary $\kap a$ and small zeta potential. The particle surface is described by $r_s(\theta) = a[1 + \eps f(\theta)]$, where $\eps \ll 1$ and $f(\theta)$ is expanded in Legendre polynomials. The perturbation to the fields is obtained via the domain perturbation technique of \citet{brenner1964}, and the correction to the Stokes flow around the deformed particle is computed analytically using a Gegenbauer streamfunction decomposition. While the results are formally valid for small deformations ($\eps \ll 1$), they should provide useful physical insight and motivate the broader question of shape-dependent electrophoretic mobility.

The remainder of this paper is organized as follows. In \S\,\ref{sec:setup}, we define the problem and state the governing equations. In \S\,\ref{sec:derivation}, we carry out the perturbation analysis and compute each contribution to the shape correction coefficient. In \S\,\ref{sec:results}, we present the resulting mobility corrections and validate the results against the exact solutions of \citet{yoon1989}. In \S\,\ref{sec:conclusions}, we summarize the key findings and discuss directions for future work. Finally, we note that this work was developed with the assistance of Claude (Anthropic), as detailed in \S\,\ref{sec:ai}.

\section{Problem setup}
\label{sec:setup}

\subsection{Geometry and physical setting}

\begin{figure}
\centering
\includegraphics[width=0.5\textwidth]{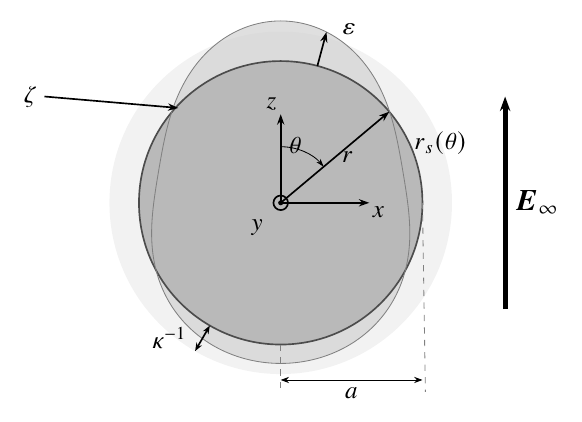}
\caption{Geometry of the problem. A nearly spherical particle with surface $r_s(\theta) = a[1 + \eps f(\theta)]$ (lighter shade) is immersed in an electrolyte with Debye length $\kap^{-1}$ (outer shaded region). The darker circle indicates the reference sphere of radius $a$. The applied electric field $\bE_\infty$ is directed along the $z$-axis. The particle carries a uniform zeta potential $\zeta$.}
\label{fig:schematic}
\end{figure}

We consider a rigid, non-conducting particle immersed in a symmetric binary electrolyte of viscosity $\mu$, permittivity $\eps_f$, and Debye length $\kap^{-1} = (\eps_f k_B T / 2 e^2 c_0)^{1/2}$, where $k_B$ is the Boltzmann constant, $T$ is the temperature, $e$ is the elementary charge, and $c_0$ is the bulk electrolyte concentration. The particle surface is described by
\begin{equation}
r_s(\theta) = a\bigl[1 + \eps f(\theta)\bigr], \qquad f(\theta) = \sum_{n=0}^{\infty} c_n P_n(\cos\theta),
\label{eq:shape}
\end{equation}
where $a$ is the radius of a reference sphere, $\eps \ll 1$ is a small dimensionless parameter characterizing the amplitude of the shape deformation, $P_n$ are Legendre polynomials, and $c_n$ are order-unity coefficients. We restrict attention to axisymmetric shapes, so that the surface depends only on the polar angle $\theta$ measured from the axis of the applied field. The particle carries a uniform zeta potential $\zeta$ on its surface, and the analysis is conducted within the Debye--H\"{u}ckel (linearized Poisson--Boltzmann) approximation, $e\zeta / k_B T \ll 1$. A uniform electric field $\bE_\infty = E_\infty \hat{\be}_z$ is applied far from the particle. The geometry is illustrated in figure~\ref{fig:schematic}.

\subsection{Governing equations}

Under the assumptions stated above (small $\zeta$, weak applied field, negligible ionic P\'{e}clet number), the electrokinetic problem decomposes into three independent sub-problems \citep{henry1931, obrien1978, ganguly2024} - (i) the equilibrium double-layer potential $\psi$, (ii) the applied-field potential $\Phi$, and (iii) the fluid velocity driven by the resulting body force. The equilibrium double-layer potential $\psi$ satisfies the linearized Poisson--Boltzmann equation
\begin{equation}
\nabla^2 \psi = \kap^2 \psi \quad \text{in} \quad r > r_s,
\label{eq:psi_gov}
\end{equation}
with $\psi = \zeta$ on $r = r_s$ and $\psi \to 0$ as $r \to \infty$. We adopt the constant surface potential condition here; an equivalent formulation specifies a constant surface charge density $\sigma$ via Gauss's law, $-\eps_f \, \partial\psi/\partial n|_{r_s} = \sigma$. For the weak-field electrophoresis problem in the linearized regime, \citet{obrien1978} showed that the mobility is independent of whether one specifies constant potential or constant charge, a point also noted by \citet{yoon1989} for spheroids and by \citet{ganguly2024} for arbitrary particle shapes. The two formulations therefore yield identical results for the shape corrections computed in this work; for a recent and detailed comparison of the constant-potential and constant-charge formulations, including the nonlinear high-potential regime where this equivalence no longer holds, see \citet{casarella2025}. We note, however, that this equivalence does not extend to diffusiophoresis, where the imposed solute gradient perturbs the equilibrium ion distribution and the surface electrostatic condition enters the mobility at leading order \citep{prieve1984, gupta2019}. Thus, while the electrophoretic results of this paper are insensitive to the choice of boundary condition, a generalization to diffusiophoresis would require specifying the surface condition explicitly.

The potential associated with the applied electric field, $\Phi$, satisfies Laplace's equation
\begin{equation}
\nabla^2 \Phi = 0 \quad \text{in} \quad r > r_s,
\label{eq:Phi_gov}
\end{equation}
with the insulating boundary condition $\partial \Phi / \partial n = 0$ on $r = r_s$ and $\Phi \to -E_\infty r \cos\theta$ as $r \to \infty$. Following \citet{brady2021} and \citet{ganguly2024}, the fluid motion is driven by an osmophoretic body force
\begin{equation}
\bb = \eps_f \kap^2 \psi \nabla\Phi,
\label{eq:bodyforce}
\end{equation}
which arises at leading order in the Debye--H\"{u}ckel limit for a symmetric electrolyte. The body force is concentrated in the diffuse double layer, where the equilibrium charge density $\rho_e = -\eps_f \kap^2 \psi$ is non-zero. We emphasise that the decomposition into $\psi$ and $\Phi$ is a consequence of the weak-field (linear) approximation and holds at arbitrary $\kap a$.

\subsection{Unified mobility expression}

The translational velocity of the particle is obtained from a reciprocal theorem framework. For a particle of arbitrary shape, the result reads
\begin{equation}
\bU = \mathbf{M}_{UF} \cdot \int_V (\DT - \I) \cdot \bb \, \dd V,
\label{eq:mobility_general}
\end{equation}
where $\mathbf{M}_{UF}$ is the hydrodynamic mobility tensor relating force to translational velocity, $\DT$ is the disturbance tensor for translation (i.e.\ $\DT \cdot \hat{U}\hat{\be}_z = \hat{\bu}$, the Stokes velocity field for the particle translating with velocity $\hat{U}\hat{\be}_z$), and $\I$ is the identity tensor. The integration is over the entire fluid volume $V$ exterior to the particle. The identity subtraction in $(\DT - \I)$ removes the direct osmophoretic force, leaving only the hydrodynamic correction due to the disturbance flow near the particle surface. Here $\hat{U}$ is an arbitrary reference speed whose value cancels in the final mobility; it serves only to define $\DT$ as a purely geometric (shape-dependent) quantity. For a sphere of radius $a$, $\mathbf{M}_{UF} = (6\pi\mu a)^{-1}\I$ and the disturbance tensor takes the form \citep{kim2005}
\begin{equation}
\DT = \frac{3a}{4r}(\I + \be_r \be_r) + \frac{a^3}{4r^3}(\I - 3\be_r \be_r),
\label{eq:DT_sphere}
\end{equation}
where $\be_r$ is the radial unit vector. An important property of (\ref{eq:DT_sphere}) is that $\DT \to \I$ as $r \to a$, so that $(\DT - \I) \cdot \bb$ vanishes on the sphere surface. This vanishing will have direct consequences for the domain correction at $O(\eps)$.

Projecting (\ref{eq:mobility_general}) onto the $z$-axis and substituting the leading-order fields on the reference sphere recovers the classical result of \citet{henry1931},
\begin{equation}
U_0 = \frac{\eps_f \zeta E_\infty}{\mu} \, f_H(\kap a),
\label{eq:henry}
\end{equation}
where $f_H(\kap a)$ is Henry's function, which interpolates monotonically from $f_H(0) = 2/3$ (the H\"{u}ckel limit) to $f_H(\infty) = 1$ (the Smoluchowski limit). Interested readers are referred to \citet{ganguly2024} for a more detailed derivation of this result.

\subsection{Perturbation structure}

To compute the $O(\eps)$ correction to the mobility, we expand all fields in powers of $\eps$: $\psi = \psi_0 + \eps \psi_1 + \cdots$, $\Phi = \Phi_0 + \eps \Phi_1 + \cdots$, $\bb = \bb_0 + \eps \bb_1 + \cdots$. The deformed particle also modifies the disturbance tensor, $\DT = \DT_0 + \eps \DT_1 + \cdots$, the mobility tensor, $\mathbf{M}_{UF} = (6\pi\mu a)^{-1}(1 + \eps \alpha_n^{\rm drag} + \cdots)\I$ (where $\alpha_n^{\rm drag}$ is the Brenner drag correction), and the integration domain, $V = V_0 + \eps\,\delta V + \cdots$. The perturbation fields $\psi_1$ and $\Phi_1$ are obtained by Taylor-expanding the boundary conditions from the deformed surface $r = r_s$ to the reference sphere $r = a$ and solving the resulting boundary-value problems on the sphere, following the domain perturbation technique of \citet{brenner1964}. The perturbation to the Stokes flow, $\hat{\bu}_1 = \DT_1 \cdot \hat{U}\hat{\be}_z$, is likewise obtained by projecting the no-slip condition onto the reference sphere.

The electrophoretic velocity at $O(\eps)$ may be written as
\begin{equation}
U = U_0 \bigl[1 + \eps \, c_n \, \sigma_n(\kap a) + O(\eps^2)\bigr],
\label{eq:U_expansion}
\end{equation}
where $\sigma_n(\kap a)$ is a shape correction coefficient that depends on the Legendre mode $n$ and the dimensionless Debye length $\kap a$. Collecting terms at $O(\eps)$ in (\ref{eq:mobility_general}) yields
\begin{equation}
\sigma_n(\kap a) = \frac{\mathcal{I}^{(\psi_1)} + \mathcal{I}^{(\Phi_1)} + \mathcal{I}^{(\hat{\bu}_1)} + \mathcal{I}^{(\rm dom)}}{\mathcal{J}_0} + \alpha_n^{\rm drag},
\label{eq:sigma_decomp}
\end{equation}
where $\mathcal{J}_0$ is the base integral that produces Henry's result (\ref{eq:henry}), and the four numerator integrals arise respectively from - (i) the perturbation to the equilibrium potential ($\psi_1$), (ii) the perturbation to the applied field ($\Phi_1$), (iii) the perturbation to the Stokes flow ($\hat{\bu}_1$), and (iv) the shift in the integration domain ($\delta V$). Each of these contributions has a distinct physical origin and $\kap a$-dependence, and they are evaluated in detail in \S\,\ref{sec:derivation}.

\section{Mathematical derivation}
\label{sec:derivation}

We now evaluate each contribution to the shape correction coefficient $\sigma_n(\kap a)$ defined in (\ref{eq:sigma_decomp}). For concreteness, we present the complete derivation for the quadrupolar mode $f(\theta) = c_2 P_2(\cos\theta)$, which is the leading shape deformation that produces a net mobility correction. Higher modes are discussed in \S\,\ref{sec:higher_modes}.

\subsection{Leading-order fields and Henry's function}
\label{sec:leading_order}

On the reference sphere, the equilibrium potential satisfying (\ref{eq:psi_gov}) with $\psi_0(a) = \zeta$ and $\psi_0 \to 0$ at infinity is
\begin{equation}
\psi_0(r) = \zeta \frac{a}{r} \, e^{-\kap(r-a)}.
\label{eq:psi0}
\end{equation}
The exponential decay on the scale $\kap^{-1}$ reflects the screening of the surface charge by the diffuse double layer, while the $a/r$ prefactor is the familiar electrostatic potential of a sphere. The applied field potential satisfying (\ref{eq:Phi_gov}) with $\partial\Phi_0/\partial r = 0$ at $r = a$ and $\Phi_0 \to -E_\infty r\cos\theta$ at infinity is
\begin{equation}
\Phi_0(r,\theta) = -E_\infty\left(r + \frac{a^3}{2r^2}\right)\cos\theta.
\label{eq:Phi0}
\end{equation}
The dipole correction $a^3/(2r^2)$ arises from the insulating boundary condition and enhances the tangential electric field at the particle surface by a factor of $3/2$ relative to the imposed field.

Substituting $\bb_0 = \eps_f\kap^2\psi_0\nabla\Phi_0$ into the $z$-component of (\ref{eq:mobility_general}) and performing the angular integration recovers the classical result $U_0 = (\eps_f\zeta E_\infty/\mu)\,f_H(\kap a)$, with Henry's function given by \citep{henry1931}
\begin{equation}
f_H(\kap a) = 1 + 2\int_1^\infty \frac{e^{-\kap a(s-1)}}{s^5}\,\dd s - 5\int_1^\infty \frac{e^{-\kap a(s-1)}}{s^7}\,\dd s,
\label{eq:fH_integral}
\end{equation}
where $s = r/a$. In the thick-double-layer limit ($\kap a \to 0$), the exponential factor approaches unity throughout the fluid and the integrals evaluate to $1/4$ and $1/6$, yielding $f_H(0) = 1 + 1/2 - 5/6 = 2/3$ (the H\"{u}ckel result). In the thin-double-layer limit ($\kap a \to \infty$), the exponential factor $e^{-\kap a(s-1)}$ confines both integrands to $s \approx 1$, where $1/s^5 \approx 1/s^7 \approx 1$, so each integral reduces to $\int_1^\infty e^{-\kap a(s-1)}\,\dd s = 1/(\kap a)$ and $f_H \to 1 + 2/(\kap a) - 5/(\kap a) = 1 - 3/(\kap a)$, recovering the Smoluchowski result $f_H = 1$ at leading order with a first correction of $-3/(\kap a)$ at finite $\kap a$.

For the perturbation analysis, it is useful to define the base integral
\begin{equation}
\mathcal{J}_0 = 2\pi a^3 \int_1^\infty \frac{e^{-\kap a(s-1)}}{s}\left(\frac{2\tilde{A}}{3} + \frac{4\tilde{B}}{3}\right)s^2\,\dd s,
\label{eq:J0}
\end{equation}
which is the volume integral producing $U_0$, with the angular kernel functions
\begin{equation}
\tilde{A}(s) = -\!\left(1-\frac{1}{s^3}\right)\!\left(1 - \frac{3}{2s} + \frac{1}{2s^3}\right)\!, \quad
\tilde{B}(s) = \left(1+\frac{1}{2s^3}\right)\!\left(1 - \frac{3}{4s} - \frac{1}{4s^3}\right)\!.
\label{eq:AB_kernel}
\end{equation}
These arise from the $z$-projection of $[(\DT_0 - \I)\cdot\bb_0]_z = \eps_f\kap^2\psi_0 E_\infty[\tilde{A}(s)\cos^2\theta + \tilde{B}(s)\sin^2\theta]$, where the factors of $\cos^2\theta$ and $\sin^2\theta$ come from projecting the body force components ($b_r\cos\theta$ and $b_\theta\sin\theta$) onto $\be_z$ through the $(\DT_0 - \I)$ kernel. Both kernel functions vanish at $s = 1$, $\tilde{A}(1) = \tilde{B}(1) = 0$, consistent with the property $(\DT_0 - \I)|_{r=a} = \mathbf{0}$ noted in \S\,\ref{sec:setup}.

\subsection{Perturbation to the equilibrium potential}
\label{sec:psi1}

The deformed surface carries the same potential $\zeta$, so the boundary condition $\psi(r_s) = \zeta$ must hold to all orders in $\eps$. Taylor-expanding from $r = r_s$ to $r = a$ at $\Order{\eps}$ gives
\begin{equation}
\psi_1(a,\theta) = -a\,c_2\,P_2(\cos\theta) \left.\frac{\partial\psi_0}{\partial r}\right|_{r=a} = \zeta\,c_2\,(1+\kap a)\,P_2(\cos\theta),
\label{eq:psi1_BC_deriv}
\end{equation}
where we have used $\partial\psi_0/\partial r|_{r=a} = -\zeta(1+\kap a)/a$ from (\ref{eq:psi0}). Physically, the boundary value of $\psi_1$ is large when $\kap a \gg 1$ because the equilibrium potential drops steeply near the surface; shifting the surface by a small amount $\eps a$ produces a perturbation of order $\eps\kap a$ relative to $\zeta$. The perturbation expansion is therefore well-ordered only when $\eps\kap a \ll 1$, i.e.\ when the surface deformation is small compared to the Debye length. We note, however, that the final result for $\sigma_2(\kap a)$ does recover the correct Smoluchowski limit $\sigma_2 \to 0$ as $\kap a \to \infty$ (\S\,\ref{sec:smoluchowski_limit}), so the shape correction vanishes precisely in the regime where the ordering constraint is most severe. Nonetheless, this constraint should not be overlooked when applying the theory at large but finite $\kap a$.

The perturbation $\psi_1$ satisfies the same linearized Poisson--Boltzmann equation, $\nabla^2\psi_1 = \kap^2\psi_1$, in $r > a$, with $\psi_1 \to 0$ at infinity. The solution is
\begin{equation}
\psi_1(r,\theta) = \zeta\,c_2\,(1+\kap a)\,\frac{k_2(\kap r)}{k_2(\kap a)}\,P_2(\cos\theta),
\label{eq:psi1_sol}
\end{equation}
where $k_n(x) = \sqrt{\pi/(2x)}\,K_{n+1/2}(x)$ is the modified spherical Bessel function of the second kind, which is the unique solution of $(\nabla^2 - \kap^2)\phi = 0$ with $P_n(\cos\theta)$ angular dependence that decays at infinity. The ratio $k_2(\kap r)/k_2(\kap a)$ can be evaluated numerically using standard library functions for $K_{n+1/2}$ (e.g.\ \texttt{scipy.special.kv}).

The contribution of $\psi_1$ to the mobility correction arises from the body force perturbation $\eps_f\kap^2\psi_1\nabla\Phi_0$, projected through the $(\DT_0 - \I)$ kernel. Since $\psi_1 \propto P_2(\cos\theta)$ and the kernel functions $\tilde{A}(s)\cos^2\theta + \tilde{B}(s)\sin^2\theta$ contain only $P_0$ and $P_2$ angular components (because $\cos^2\theta = (2P_2+1)/3$ and $\sin^2\theta = 2(1-P_2)/3$), the angular integral reduces by Legendre orthogonality to
\begin{equation}
\mathcal{I}^{(\psi_1)} = 2\pi a^3\,(1+\kap a)\int_1^\infty \frac{k_2(\kap a s)}{k_2(\kap a)}\,\frac{4(\tilde{A}-\tilde{B})}{15}\,s^2\,\dd s.
\label{eq:Ipsi1}
\end{equation}
The angular projection coefficient $4(\tilde{A} - \tilde{B})/15$ arises from $(2/3)\int_0^\pi P_2^2\,\sin\theta\,\dd\theta = 4/15$. The radial integral is evaluated numerically using Gauss quadrature, exploiting the exponential decay of $k_2(\kap a s)$ at large $s$.

This result has an important corollary. For any shape mode $n \neq 0, 2$, the angular integral $\int_0^\pi P_n\,[\tilde{A}\cos^2\theta + \tilde{B}\sin^2\theta]\,\sin\theta\,\dd\theta$ vanishes by orthogonality, since the integrand contains only $P_0$ and $P_2$ angular components. Therefore, \emph{the $\psi_1$ contribution to $\sigma_n$ vanishes identically for all $n \geq 3$} (and for $n = 1$, which corresponds to a rigid translation and is not a genuine shape deformation).

\subsection{Perturbation to the applied field}
\label{sec:Phi1}

The insulating boundary condition on the deformed surface requires more care than the Dirichlet condition for $\psi$, because the outward unit normal to $r = r_s(\theta)$ is tilted relative to $\be_r$ at $\Order{\eps}$. For $r_s = a[1 + \eps f(\theta)]$, the outward normal is
\begin{equation}
\bn = \be_r - \eps\,f'(\theta)\,\be_\theta + \Order{\eps^2},
\label{eq:normal}
\end{equation}
where $f' = \dd f/\dd\theta$. Expanding the condition $\bn\cdot\nabla\Phi|_{r_s} = 0$ to $\Order{\eps}$ yields two contributions - one from Taylor-expanding $\partial\Phi_0/\partial r$ from $r_s$ to $r = a$, and one from the tilted normal acting on $\Phi_0$. Together they give
\begin{equation}
\left.\frac{\partial\Phi_1}{\partial r}\right|_{r=a} = -a\,f(\theta)\left.\frac{\partial^2\Phi_0}{\partial r^2}\right|_{r=a} + \frac{f'(\theta)}{a}\left.\frac{\partial\Phi_0}{\partial\theta}\right|_{r=a}.
\label{eq:Phi1_BC_general}
\end{equation}
From (\ref{eq:Phi0}), the required derivatives at $r = a$ are
\begin{equation}
\left.\frac{\partial^2\Phi_0}{\partial r^2}\right|_{r=a} = -\frac{3E_\infty\cos\theta}{a}, \qquad
\left.\frac{\partial\Phi_0}{\partial\theta}\right|_{r=a} = \frac{3}{2}\,E_\infty a\sin\theta.
\label{eq:Phi0_derivs}
\end{equation}
For $f(\theta) = c_2 P_2(\cos\theta)$, we have $f'(\theta) = -3c_2\sin\theta\cos\theta$, and (\ref{eq:Phi1_BC_general}) becomes
\begin{equation}
\left.\frac{\partial\Phi_1}{\partial r}\right|_{r=a} = 3c_2 E_\infty P_2\cos\theta - \frac{9}{2}\,c_2 E_\infty\sin^2\!\theta\cos\theta.
\label{eq:Phi1_BC_raw}
\end{equation}
The first term is $3c_2 E_\infty P_2\cos\theta$, which arises from the curvature of $\Phi_0$ in the radial direction; the second, $-(9/2)c_2 E_\infty\sin^2\theta\cos\theta$, arises from the normal tilt. Expanding in Legendre polynomials using the identities $P_2\cos\theta = (2P_1 + 3P_3)/5$ and $\sin^2\!\theta\cos\theta = (2P_1 - 2P_3)/5$ gives
\begin{equation}
\left.\frac{\partial\Phi_1}{\partial r}\right|_{r=a} = c_2\,E_\infty\left(-\frac{3}{5}\cos\theta + \frac{18}{5}\,P_3\right).
\label{eq:Phi1_BC_expanded}
\end{equation}
The deformation thus excites $P_1$ (dipole) and $P_3$ (octupole) harmonics in the perturbed applied field, even though the shape perturbation itself is purely $P_2$.

The solution of Laplace's equation decaying at infinity with the Neumann condition (\ref{eq:Phi1_BC_expanded}) is
\begin{equation}
\Phi_1(r,\theta) = c_2\,E_\infty\,a\left[\frac{3}{10}\left(\frac{a}{r}\right)^{\!2}\cos\theta - \frac{9}{10}\left(\frac{a}{r}\right)^{\!4} P_3(\cos\theta)\right].
\label{eq:Phi1_sol}
\end{equation}
The coefficients are obtained by matching (\ref{eq:Phi1_BC_expanded}) mode by mode. The exterior harmonic $(a/r)^2\cos\theta$ has radial derivative $-2\cos\theta/a$ at $r = a$ (from the $P_1$ mode), yielding the factor $3/10$, while $(a/r)^4 P_3$ has derivative $-4P_3/a$ (from the $P_3$ mode), yielding $-9/10$.

The contribution of $\Phi_1$ to the mobility correction is
\begin{equation}
\mathcal{I}^{(\Phi_1)} = \eps_f\kap^2\int_{V_0}\left[(\DT_0 - \I)\cdot(\psi_0\nabla\Phi_1)\right]_z\,\dd V.
\label{eq:IPhi1}
\end{equation}
Unlike the $\psi_1$ integral, the angular structure of $\nabla\Phi_1$ involves both $P_1$ and $P_3$ harmonics with $r$-dependent gradient components, so the angular integral at each radial station does not simplify to a closed form. We evaluate (\ref{eq:IPhi1}) by performing the angular integration numerically via Gauss--Legendre quadrature at each $s$, followed by Gauss quadrature in the radial direction, exploiting the exponential decay of $\psi_0$ at large $s$.

\subsection{Domain and drag corrections}
\label{sec:domain_drag}

\subsubsection{Domain correction}

The shift of the integration domain from the reference sphere ($r > a$) to the actual domain ($r > r_s$) produces, at $\Order{\eps}$, a surface integral over the shell between $r = a$ and $r = r_s$,
\begin{equation}
\mathcal{I}^{(\rm dom)} = -\eps_f\kap^2\oint_{r=a}a\,f(\theta)\left[(\DT_0 - \I)\cdot\bb_0\right]_z\Big|_{r=a}\,\dd S.
\label{eq:Idom_formula}
\end{equation}
However, the key property of the $(\DT - \I)$ formulation is that $(\DT_0 - \I)\cdot\bb_0$ vanishes at $r = a$, because $\DT_0|_{r=a} = \I$. Therefore,
\begin{equation}
\mathcal{I}^{(\rm dom)} = 0.
\label{eq:Idom_zero}
\end{equation}
This vanishing simplifies the perturbation analysis considerably.

\subsubsection{Drag correction}

The shape deformation modifies the Stokes resistance of the particle, entering through the mobility tensor $\mathbf{M}_{UF}$ in (\ref{eq:mobility_general}). For a nearly spherical particle with surface $r_s = a[1 + \eps \sum_k a_k P_k(\cos\theta)]$, \citet{brenner1964} showed (his equation~1.7; crediting \citealt{paynepell1960}) that the Stokes drag force for translation along the symmetry axis is
\begin{equation}
F = 6\pi\mu a U\left[1 + \eps\!\left(a_0 - \frac{1}{5}\,a_2\right) + \Order{\eps^2}\right].
\label{eq:brenner_drag}
\end{equation}
Only the $k = 0$ and $k = 2$ modes contribute at $\Order{\eps}$. For a pure $P_2$ deformation ($a_0 = 0$, $a_2 = c_2$), the drag \emph{decreases} for a prolate shape ($c_2 > 0$), which is physically intuitive. A particle elongated along its direction of motion is more streamlined. Inverting (\ref{eq:brenner_drag}) to obtain the mobility,
\begin{equation}
\alpha_2^{\rm drag} = +\frac{1}{5}.
\label{eq:alpha_drag}
\end{equation}

For higher modes $n \geq 3$, Brenner showed that $\alpha_n^{(F)} = 0$. The Stokes drag is unaffected at $\Order{\eps}$ by shape perturbations of order $P_n$ with $n \geq 3$, because such modes do not couple to the Stokeslet (dipole) far-field component that determines the net force \citep{brenner1964, kim2005}. As a result, $\alpha_n^{\rm drag} = 0$ for $n \geq 3$.

\subsection{Perturbation to the Stokes disturbance flow}
\label{sec:u1}

The perturbation to the disturbance tensor, $\DT_1$, represents the change in the Stokes velocity field around the deformed particle. Since $\DT_1\cdot\hat{U}\be_z = \bhat{u}_1$, the first-order velocity perturbation, this contribution requires solving the Stokes equations for $\bhat{u}_1$ and computing
\begin{equation}
\mathcal{I}^{(\bhat{u}_1)} = \eps_f\kap^2\int_{V_0}\psi_0\,(\nabla\Phi_0\cdot\bhat{u}_1)\,\dd V.
\label{eq:Iu1}
\end{equation}
Recall that $\hat{U}\be_z$ is simply an arbitrary translation velocity in the $z$-direction, as discussed below (\ref{eq:mobility_general}). A subtle but important point is that this integral involves $\DT_1$ only (not $\DT_1 - \I$), since the identity tensor is unchanged at $\Order{\eps}$. The integrand therefore need not vanish at $r = a$, in contrast to the base integrand.

\subsubsection{Boundary conditions}

The reference disturbance field $\bhat{u}_0$ satisfies the no-slip condition $\bhat{u}_0(a) = \hat{U}\be_z$. On the deformed surface, we require $\bhat{u}(r_s) = \hat{U}\be_z$. Taylor-expanding at $\Order{\eps}$ -
\begin{equation}
\bhat{u}_1(a,\theta) = -a\,c_2\,P_2(\cos\theta)\left.\frac{\partial\bhat{u}_0}{\partial r}\right|_{r=a}.
\label{eq:u1_BC_vector}
\end{equation}
The components of the reference velocity field are $\hat{u}_{0,r} = \hat{U}(3/(2s) - 1/(2s^3))\cos\theta$ and $\hat{u}_{0,\theta} = -\hat{U}(3/(4s) + 1/(4s^3))\sin\theta$, where $s = r/a$. Evaluating the radial derivatives at $s = 1$,
\begin{equation}
\left.\frac{\partial\hat{u}_{0,r}}{\partial r}\right|_{r=a} = 0, \qquad
\left.\frac{\partial\hat{u}_{0,\theta}}{\partial r}\right|_{r=a} = \frac{3\hat{U}}{2a}\sin\theta.
\label{eq:u0_derivs}
\end{equation}
The vanishing of $\partial\hat{u}_{0,r}/\partial r|_{r=a}$ follows from the structure of the Stokes solution. The radial velocity $\hat{u}_{0,r}(a) = \hat{U}\cos\theta$ is already set by the kinematic boundary condition, and the combination of no-slip and incompressibility forces $\partial\hat{u}_{0,r}/\partial r|_{r=a} = 0$. Therefore,
\begin{equation}
\hat{u}_{1,r}(a,\theta) = 0, \qquad \hat{u}_{1,\theta}(a,\theta) = -\frac{3\hat{U}}{2}\,c_2\,P_2(\cos\theta)\sin\theta.
\label{eq:u1_BC}
\end{equation}
At infinity, $\bhat{u}_1 \to \mathbf{0}$.

\subsubsection{Gegenbauer streamfunction decomposition}
\label{sec:gegenbauer}

For axisymmetric Stokes flow, the velocity field can be expressed in terms of a Stokes streamfunction $\Psi_1(r,\theta)$,
\begin{equation}
\hat{u}_{1,r} = \frac{1}{r^2\sin\theta}\frac{\partial\Psi_1}{\partial\theta}, \qquad
\hat{u}_{1,\theta} = -\frac{1}{r\sin\theta}\frac{\partial\Psi_1}{\partial r},
\label{eq:stokes_stream}
\end{equation}
and the Stokes equations reduce to the biharmonic equation $E^4\Psi_1 = 0$, where $E^2 = \partial^2/\partial r^2 + (\sin\theta/r^2)\,\partial/\partial\theta\,(1/\sin\theta\,\partial/\partial\theta)$ is the Stokes stream operator \citep{kim2005}. The general solution is expanded in Gegenbauer functions $C_m^{-1/2}(\cos\theta)$,
\begin{equation}
\Psi_1 = \sum_m h_m(r)\,C_m^{-1/2}(\cos\theta),
\end{equation}
where each radial function $h_m(r)$ satisfies the fourth-order ordinary differential equation
\begin{equation}
\left[\frac{\dd^2}{\dd r^2} - \frac{m(m-1)}{r^2}\right]^2 h_m = 0.
\label{eq:biharmonic_radial}
\end{equation}

\emph{Angular decomposition.} The tangential boundary condition (\ref{eq:u1_BC}) involves $P_2(\cos\theta)\sin\theta$, which must be expressed in terms of the angular basis associated with the Gegenbauer modes. Writing $P_2 = (3\cos^2\theta - 1)/2$ and using the decomposition
\begin{equation}
3\cos^2\theta - 1 = -\frac{2}{5} + \frac{3}{5}\,(5\cos^2\theta - 1),
\label{eq:cos2_decomp}
\end{equation}
the tangential boundary condition becomes
\begin{equation}
\hat{u}_{1,\theta}(a) = \frac{3\hat{U}c_2}{10}\sin\theta - \frac{9\hat{U}c_2}{20}(5\cos^2\theta - 1)\sin\theta.
\label{eq:u1theta_decomp}
\end{equation}
The first term (proportional to $\sin\theta$) drives a Gegenbauer mode $m = 2$, for which $C_2^{-1/2} = \sin^2\!\theta/2$. The second term (proportional to $(5\cos^2\!\theta - 1)\sin\theta$) drives a mode $m = 4$, for which $C_4^{-1/2} = \sin^2\!\theta\,(5\cos^2\!\theta - 1)/8$. No other Gegenbauer modes are excited.

\emph{Mode $m = 2$: Stokeslet and potential dipole.} The radial ODE (\ref{eq:biharmonic_radial}) with $m = 2$ is $[d^2/dr^2 - 2/r^2]^2 h_2 = 0$, whose four independent solutions are $r^2$, $r^{-1}$, $r^4$, and $r$. The solutions that produce velocity fields decaying at infinity are $r^{-1}$ (source dipole, $\bu \sim r^{-3}$) and $r$ (Stokeslet, $\bu \sim r^{-1}$). Applying the boundary conditions $h_2(a) = 0$ (from $\hat{u}_{1,r}(a) = 0$) and $h_2'(a) = -3\hat{U}a c_2/5$ (from the $\sin\theta$ coefficient in (\ref{eq:u1theta_decomp}), accounting for the Gegenbauer normalization and the minus sign in (\ref{eq:stokes_stream})), we obtain
\begin{equation}
h_2(r) = \frac{3\hat{U}a\,c_2}{10}\left(\frac{a^2}{r} - r\right).
\label{eq:h2}
\end{equation}

\emph{Mode $m = 4$: octupole and hexadecapole.} With $m = 4$, the ODE $[d^2/dr^2 - 12/r^2]^2 h_4 = 0$ has decaying solutions $r^{-3}$ and $r^{-1}$. The boundary conditions $h_4(a) = 0$ and $h_4'(a) = 18\hat{U}a\,c_2/5$ (from the $(5\cos^2\theta - 1)\sin\theta$ coefficient) yield
\begin{equation}
h_4(r) = \frac{9\hat{U}a^3 c_2}{5}\left(\frac{1}{r} - \frac{a^2}{r^3}\right).
\label{eq:h4}
\end{equation}

\subsubsection{Velocity field}

Differentiating the streamfunction yields the velocity components. In terms of $s = r/a$ and dividing by $\hat{U}c_2$,
\begin{align}
\frac{\hat{u}_{1,r}}{\hat{U}c_2} &= \frac{3}{10}\!\left(\frac{1}{s^3} - \frac{1}{s}\right)\cos\theta + \frac{9}{5}\!\left(\frac{1}{s^3} - \frac{1}{s^5}\right)P_3(\cos\theta), \label{eq:u1r}\\[6pt]
\frac{\hat{u}_{1,\theta}}{\hat{U}c_2} &= \frac{3}{20}\!\left(\frac{1}{s^3} + \frac{1}{s}\right)\sin\theta + \frac{9}{40}\!\left(\frac{1}{s^3} - \frac{3}{s^5}\right)(5\cos^2\!\theta - 1)\sin\theta. \label{eq:u1th}
\end{align}
The $m = 2$ mode (first terms) decays as $1/s$ at large $s$, characteristic of a Stokeslet; the $m = 4$ mode (second terms) decays as $1/s^3$, characteristic of an octupole. One can verify that (\ref{eq:u1r})--(\ref{eq:u1th}) satisfy $\hat{u}_{1,r}(1) = 0$ and $\hat{u}_{1,\theta}(1) = -(3/2)P_2\sin\theta$.

The $\bhat{u}_1$ contribution (\ref{eq:Iu1}) is evaluated by numerical integration over both angle and radius, with the integrand given by $\psi_0(\nabla\Phi_0\cdot\bhat{u}_1)$ where the analytical expressions (\ref{eq:u1r})--(\ref{eq:u1th}) are used for $\bhat{u}_1$ and the leading-order fields (\ref{eq:psi0})--(\ref{eq:Phi0}) for $\psi_0$ and $\nabla\Phi_0$.

\subsection{Assembly and numerical evaluation}
\label{sec:assembly}

Collecting the results from \S\S\,\ref{sec:psi1}--\ref{sec:u1}, the shape correction coefficient for the $P_2$ mode is
\begin{equation}
\sigma_2(\kap a) = \frac{\mathcal{I}^{(\psi_1)} + \mathcal{I}^{(\Phi_1)} + \mathcal{I}^{(\bhat{u}_1)}}{\mathcal{J}_0} + \frac{1}{5},
\label{eq:sigma2_assembly}
\end{equation}
where the domain correction is zero and the drag correction is $\alpha_2^{\rm drag} = +1/5$ (cf.\ (\ref{eq:alpha_drag})). The electrophoretic velocity of a nearly spherical particle with surface $r_s = a[1 + \eps c_2 P_2(\cos\theta)]$ is therefore
\begin{equation}
U = \frac{\eps_f\zeta E_\infty}{\mu}\,f_H(\kap a)\bigl[1 + \eps\,c_2\,\sigma_2(\kap a) + \Order{\eps^2}\bigr].
\label{eq:U_final}
\end{equation}

All integrals in (\ref{eq:sigma2_assembly}) are evaluated numerically. The radial integrals extend from $s = 1$ to $s = \infty$ but are dominated by the region $1 < s \lesssim 1 + \text{few}/(\kap a)$ due to the exponential decay of $\psi_0$ and $k_2(\kap a s)$. We use Gauss quadrature with a change of variables suited to the exponential decay for the radial integration, and Gauss--Legendre quadrature for the angular integrals over $\theta \in [0,\pi]$. The modified spherical Bessel functions are computed via $k_n(x) = \sqrt{\pi/(2x)}\,K_{n+1/2}(x)$ using standard library routines.

\subsection{Higher modes: $n \geq 3$}
\label{sec:higher_modes}

The analysis above was carried out for the $n = 2$ mode. For higher-order shape perturbations $f(\theta) = c_n P_n(\cos\theta)$ with $n \geq 3$, every term in the assembly formula (\ref{eq:sigma_decomp}) vanishes identically:
\begin{enumerate}
\item The drag correction vanishes: $\alpha_n^{\rm drag} = 0$ \citep{brenner1964}.
\item The $\psi_1$ contribution vanishes by Legendre orthogonality (\S\,\ref{sec:psi1}).
\item The $\Phi_1$ contribution vanishes: $\Phi_1$ contains only $P_1$ and $P_{n-1}, P_{n+1}$ harmonics generated by the product $P_n\cos\theta$ and the normal tilt correction, but its projection onto the Henry-type integrand $\psi_0\,\nabla\Phi_0$, which has a purely $P_1$ angular structure, produces no net $P_1$ component for any $n \geq 3$.
\item The $\bhat{u}_1$ contribution vanishes: the Gegenbauer decomposition of $P_n\sin\theta$ produces Stokes-flow modes $m = n-1$ and $m = n+1$, neither of which projects onto $P_1$ when combined with the body force $\psi_0\,\nabla\Phi_0$.
\item The domain correction is zero, as for $n = 2$.
\end{enumerate}
Consequently, $\sigma_n \equiv 0$ for all $n \geq 3$: the quadrupolar mode is the only Legendre mode that contributes to the electrophoretic mobility at $\Order{\eps}$. A physical interpretation is given in \S\,\ref{sec:silencing}.

\subsection{Asymptotic limits}
\label{sec:asymptotics}

\subsubsection{H\"{u}ckel limit: $\kap a \to 0$}

In the thick-double-layer limit, the body force $\eps_f\kap^2\psi_0\nabla\Phi_0$ is spread over a volume of order $(\kap^{-1})^3 \gg a^3$, so the $\Order{\eps a}$ surface deformation has a negligible effect on the volume integrals. More precisely, the perturbation integrals $\mathcal{I}^{(\psi_1)}$, $\mathcal{I}^{(\Phi_1)}$, and $\mathcal{I}^{(\bhat{u}_1)}$ all vanish relative to $\mathcal{J}_0$ as $\kap a \to 0$. The only surviving contribution is the drag correction, giving
\begin{equation}
\sigma_2(0) = +\frac{1}{5}.
\label{eq:sigma2_huckel}
\end{equation}
This result has a clear physical interpretation: in the thick-double-layer limit, the electrostatic driving force is insensitive to particle shape, and the sole effect of deformation is to change the Stokes drag. A prolate particle ($c_2 > 0$) aligned with the field experiences less drag than a sphere of the same volume, and therefore moves faster, $\sigma_2(0) > 0$.

\subsubsection{Smoluchowski limit: $\kap a \to \infty$}
\label{sec:smoluchowski_limit}

At large $\kap a$, the volume integrals are dominated by a thin boundary layer of thickness $\Order{(\kap a)^{-1}}$ near $s = 1$. To systematically parse the competing electrostatic and hydrodynamic effects in this narrow region we have to analyze the asymptotic behavior of the three integrals: $\mathcal{J}_0$, $\mathcal{I}^{(\psi_1)}$, and $\mathcal{I}^{(\bhat{u}_1)}$. 

We first examine the unperturbed base integral $\mathcal{J}_0$. This integral features the $(\DT_0 - \I)$ kernel, which suppresses contributions near the surface by vanishing linearly as $\Order{s-1}$. This linear vanishing supplies one factor of $(\kap a)^{-1}$, and the scaled differential element supplies another, meaning the overall base integral $\mathcal{J}_0$ scales as $\Order{(\kap a)^{-2}}$.

In this region, the perturbation integrals $\mathcal{I}^{(\psi_1)}$ and $\mathcal{I}^{(\bhat{u}_1)}$ both grow as $\Order{\kap a}$ relative to $\mathcal{J}_0$, but with opposite signs,
\begin{equation}
\frac{\mathcal{I}^{(\psi_1)}}{\mathcal{J}_0} \sim -\frac{\kap a}{5}, \qquad \frac{\mathcal{I}^{(\bhat{u}_1)}}{\mathcal{J}_0} \sim +\frac{\kap a}{5} \qquad (\kap a \to \infty).
\label{eq:large_ka_growth}
\end{equation}

The prefactor $1/5$ can be exhibited explicitly. Near the surface ($s \to 1$) the base kernels (\ref{eq:AB_kernel}) behave as $\tilde{B} \sim \tfrac{9}{4}(s-1)$ with $\tilde{A} = \Order{(s-1)^2}$ subdominant, so the $\psi_1$ integrand (\ref{eq:Ipsi1}) and the base integrand of $\mathcal{J}_0$ (\ref{eq:J0}) share the same radial weight $\propto e^{-\kap a(s-1)}(s-1)$, which cancels in the ratio. What remains is the angular factor $\tfrac{4}{15}(\tilde{A} - \tilde{B}) \to -\tfrac{4}{15}\tilde{B}$ over $\tfrac{2}{3}\tilde{A} + \tfrac{4}{3}\tilde{B} \to \tfrac{4}{3}\tilde{B}$, i.e.\ $\tfrac{4}{15}\big/\tfrac{4}{3} = 1/5$, and the boundary amplification $(1 + \kap a)$ from (\ref{eq:psi1_BC_deriv}) supplies the factor of $\kap a$, giving $\mathcal{I}^{(\psi_1)}/\mathcal{J}_0 \sim -\kap a/5$. The $\bhat{u}_1$ integral shares this angular ratio but draws its $\kap a$ from a different source. Its integrand $\psi_0\,(\nabla\Phi_0\cdot\bhat{u}_1)$ carries no $(\DT_0 - \I)$ kernel and does not vanish at $s = 1$, where $\hat{u}_{1,\theta}(1) = -\tfrac{3}{2}c_2\hat{U}P_2\sin\theta$ (\ref{eq:u1_BC}) and $r^{-1}\partial_\theta\Phi_0|_{s=1} = \tfrac{3}{2}E_\infty\sin\theta$ (\ref{eq:Phi0}) give $\nabla\Phi_0\cdot\bhat{u}_1|_{s=1} = -\tfrac{9}{4}E_\infty\hat{U}c_2 P_2\sin^2\theta$; with $\int_0^\pi P_2\sin^2\theta\,\sin\theta\,\dd\theta = -\tfrac{4}{15}$ the angular ratio is again $1/5$. Lacking the kernel, its radial weight $\int e^{-\kap a(s-1)}\,\dd s \sim (\kap a)^{-1}$ carries one fewer power of $(\kap a)^{-1}$ than that of $\mathcal{J}_0$, so $\mathcal{I}^{(\bhat{u}_1)}/\mathcal{J}_0 \sim +\kap a/5$. The two thus grow equally and oppositely: $\psi_1$ through the amplified boundary value, $\bhat{u}_1$ through the lost kernel.

These two $\Order{\kap a}$ contributions nearly cancel at leading order, reflecting a physical coupling: when the Debye layer is thin, the charge redistribution ($\psi_1$) and the modified hydrodynamic response ($\bhat{u}_1$) are tightly linked. After cancellation, the $\psi_1$ and $\bhat{u}_1$ integrals together contribute an $\Order{1}$ remainder that must be determined numerically. The $\Phi_1$ contribution, however, can be evaluated analytically in this limit. Since $\Phi_1$ involves only inverse powers of $r$ (cf.\ (\ref{eq:Phi1_sol})) while $\psi_0 \sim e^{-\kap(r-a)}$ concentrates in a thin layer, the integral (\ref{eq:IPhi1}) localises to $s \approx 1$ where the $(\DT_0 - \I)$ kernel is approximately linear in $(s-1)$. The product $\psi_0(s)\,\nabla\Phi_1(s)$ is slowly varying in $(s-1)$ compared to the exponential decay, and the radial integration collapses to
\begin{equation}
\frac{\mathcal{I}^{(\Phi_1)}}{\mathcal{J}_0} \;\longrightarrow\; -\frac{1}{5} \qquad (\kap a \to \infty).
\label{eq:IPhi1_limit}
\end{equation}
This finite limit arises from the $P_1$ harmonic of $\Phi_1$ (the $(a/r)^2\cos\theta$ term with coefficient $3/10$ in (\ref{eq:Phi1_sol})), which is the only mode that survives the angular projection onto Henry's integrand. The $P_3$ component of $\Phi_1$ is orthogonal to $\cos\theta$ and does not contribute.

The shape correction coefficient in the assembly formula (\ref{eq:sigma2_assembly}) is therefore
\begin{equation}
\sigma_2(\kap a \to \infty) = \underbrace{0}_{\psi_1 + \bhat{u}_1\;(\text{cancels})} + \underbrace{\left(-\frac{1}{5}\right)}_{\Phi_1} + \underbrace{\frac{1}{5}}_{\rm drag} {}= 0.
\label{eq:sigma2_largeKa}
\end{equation}
Here the first term denotes the $\psi_1 + \bhat{u}_1$ contribution: although the two integrals are each individually $\Order{\kap a}$, their growths cancel to leave an $\Order{1}$ quantity, and numerical evaluation shows that this remainder itself vanishes in the limit. We therefore write it explicitly as zero rather than as an $\Order{1}$ residual to avoid any ambiguity. The $\Phi_1$ and drag contributions cancel exactly, so that $\sigma_2 \to 0$. The volume-integral formulation therefore recovers the Morrison--Teubner result $\sigma_n \to 0$ as $\kap a \to \infty$ in a self-consistent manner. Physically, in the thin-double-layer limit the slip velocity $\mathbf{v}_s = -(\eps_f\zeta/\mu)\bE_t|_{r_s}$ is shape-independent and drives $\bU = (\eps_f\zeta/\mu)\bE_\infty$ regardless of particle geometry \citep{morrison1970, teubner1982}. The cancellation in (\ref{eq:sigma2_largeKa}) is the perturbative expression of this theorem.

It is worth clarifying in what sense the $\kap a \to \infty$ behaviour is controlled, since the $\Order{\eps}$ expansion is well-ordered only when $\eps\kap a \ll 1$ (\S\,\ref{sec:psi1}). For fixed $\eps$ the series is asymptotic in $\kap a$ up to $\kap a \sim \Order{1/\eps}$, so the correction is quantitatively reliable at $\kap a = \Order{1}$, where it is also largest. The vanishing $\sigma_2 \to 0$ is not an artefact of subtracting two large integrals: it is fixed by the exact Morrison--Teubner theorem, which guarantees shape independence at vanishing double-layer thickness for \emph{any} geometry. The cancellation we observe, the analytic offset of $\mathcal{I}^{(\Phi_1)}$ against the drag together with the vanishing $\psi_1 + \bhat{u}_1$ remainder, is simply the $\Order{\eps}$ manifestation of that theorem; the charge redistribution and the modified hydrodynamic response are physically locked together in the thin-layer limit, so their $\Order{\kap a}$ growths must cancel. The limit is thus controlled in the iterated sense $\eps \to 0$ then $\kap a \to \infty$, and the agreement with \citet{yoon1989} across the full range of $\kap a$ (\S\,\ref{sec:yk_numerical}) confirms that this behaviour persists well beyond the naive bound $\eps\kap a \ll 1$.

\section{Results}
\label{sec:results}

\subsection{Shape correction coefficient $\sigma_2(\kap a)$}

\begin{figure}
\centering
\includegraphics[width=0.65\textwidth]{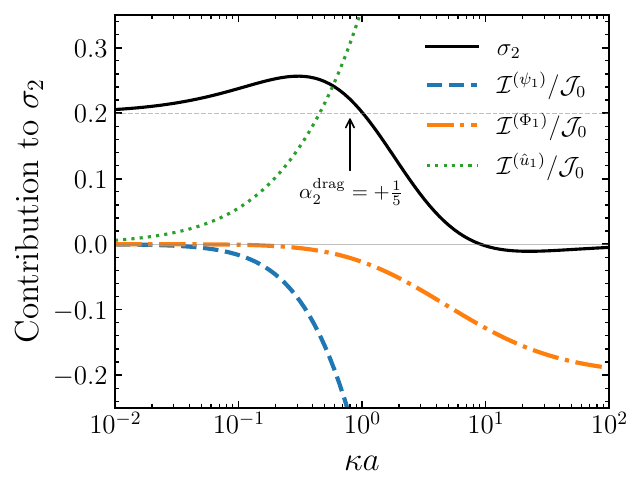}
\caption{The shape correction coefficient $\sigma_2(\kap a)$ for the $P_2$ mode and its decomposition into individual contributions: the perturbed equilibrium potential $\mathcal{I}^{(\psi_1)}/\mathcal{J}_0$, the perturbed applied field $\mathcal{I}^{(\Phi_1)}/\mathcal{J}_0$, the perturbed Stokes flow $\mathcal{I}^{(\bhat{u}_1)}/\mathcal{J}_0$, and the Brenner drag correction $\alpha_2^{\rm drag} = +1/5$ (horizontal grey dashed line). At small $\kap a$, only the drag contributes; at large $\kap a$, the $\psi_1$ and $\bhat{u}_1$ contributions nearly cancel and the $\Phi_1$ residual exactly offsets the drag. As shown in \S\,\ref{sec:higher_modes} and \S\,\ref{sec:silencing}, $\sigma_n \equiv 0$ for all $n \geq 3$, so no analogous curves exist for higher modes.}
\label{fig:sigma2}
\end{figure}

Figure~\ref{fig:sigma2} shows the shape correction coefficient $\sigma_2(\kap a)$ computed from (\ref{eq:sigma2_assembly}), together with the individual contributions from the perturbed equilibrium potential ($\mathcal{I}^{(\psi_1)}/\mathcal{J}_0$), the perturbed applied field ($\mathcal{I}^{(\Phi_1)}/\mathcal{J}_0$), the perturbed Stokes flow ($\mathcal{I}^{(\bhat{u}_1)}/\mathcal{J}_0$), and the Brenner drag correction ($\alpha_2^{\rm drag} = 1/5$). We emphasise that, as shown in \S\,\ref{sec:higher_modes} and \S\,\ref{sec:silencing}, all higher-order shape correction coefficients vanish identically, $\sigma_n \equiv 0$ for $n \geq 3$, so $\sigma_2$ is the \emph{only} non-trivial coefficient in the Legendre expansion of the $\Order{\eps}$ mobility correction.

At small $\kap a$, the only contribution is the drag correction, giving $\sigma_2 \to 1/5$ as expected from the H\"{u}ckel limit (\ref{eq:sigma2_huckel}). As $\kap a$ increases, the perturbation integrals grow: $\mathcal{I}^{(\psi_1)}$ becomes increasingly negative while $\mathcal{I}^{(\bhat{u}_1)}$ becomes positive, reflecting the opposing effects of charge redistribution and modified hydrodynamic response near the deformed surface. These two contributions grow as $\Order{\kap a}$ (cf.\ (\ref{eq:large_ka_growth})) but nearly cancel, so the total $\sigma_2$ remains $\Order{1}$. The $\Phi_1$ contribution is smaller in magnitude but approaches the finite limit $-1/5$ at large $\kap a$, where it exactly cancels the drag correction.

The total correction $\sigma_2$ is positive for $\kap a \lesssim 8$, indicating that a prolate particle ($c_2 > 0$) aligned with the field moves faster than a sphere. The correction has a maximum of approximately $0.25$ near $\kap a \approx 0.5$, then decreases through zero near $\kap a \approx 8$ and approaches zero from below as $\kap a \to \infty$, consistent with the Morrison--Teubner theorem (\ref{eq:sigma2_largeKa}).

The non-monotonic features of $\sigma_2(\kap a)$ can be read directly from this decomposition. At small $\kap a$ only the constant drag term ($+1/5$) survives and the three perturbation integrals switch on linearly in $\kap a$. Around $\kap a \approx 0.5$ the positive hydrodynamic contribution $\mathcal{I}^{(\bhat{u}_1)}/\mathcal{J}_0$ rises slightly faster than the negative charge-redistribution term $\mathcal{I}^{(\psi_1)}/\mathcal{J}_0$, while the applied-field term $\mathcal{I}^{(\Phi_1)}/\mathcal{J}_0$ has only just begun to depart from zero; the net effect lifts $\sigma_2$ to a maximum modestly \emph{above} the H\"uckel value of $1/5$, the overshoot visible in figure~\ref{fig:sigma2}. As $\kap a$ increases further, the applied-field term saturates at $\mathcal{I}^{(\Phi_1)}/\mathcal{J}_0 \to -1/5$ and offsets the drag, so that $\sigma_2$ is then governed entirely by the $\psi_1 + \bhat{u}_1$ remainder. That remainder approaches zero \emph{from below}, which produces the zero crossing near $\kap a \approx 8$ and the small negative excursion beyond it: as the thin-double-layer limit is approached, the charge-redistribution penalty transiently dominates the residual hydrodynamic gain before both decay. The negative $\sigma_2$ is admittedly counter-intuitive, since it implies a prolate particle ($c_2 > 0$) migrating marginally \emph{slower} than Henry's prediction, but it is a small $\Order{\eps}$ effect that must decay to zero as $\kap a \to \infty$ by the Morrison--Teubner theorem. The shape correction therefore changes sign once before vanishing, rather than relaxing monotonically to zero.

\subsection{Comparison with Yoon \& Kim (1989)}
\label{sec:yk_numerical}

A natural validation of the perturbation theory is comparison with the exact electrophoretic mobility of a spheroidal particle, computed by \citet{yoon1989} using spheroidal wavefunctions.

\subsubsection{Spheroid geometry and $P_2$ mapping}

A spheroid (ellipsoid of revolution) is defined in Cartesian coordinates by
\begin{equation}
\frac{x^2 + y^2}{b^2} + \frac{z^2}{c^2} = 1,
\label{eq:spheroid_cartesian}
\end{equation}
where $c$ is the polar semi-axis (along $z$) and $b$ the equatorial semi-axis. The particle is prolate when $c > b$ and oblate when $c < b$. In spherical coordinates $(r,\theta)$ centred at the origin with $z = r\cos\theta$, the surface of the spheroid is
\begin{equation}
r_s(\theta) = \frac{cb}{\sqrt{c^2\cos^2\!\theta + b^2\sin^2\!\theta}}\,.
\label{eq:spheroid_surface}
\end{equation}
Taking the equal-volume sphere $a = (cb^2)^{1/3}$ as the reference and writing $c/b = 1 + \delta$ with $\delta \ll 1$, we expand $r_s/a$ in Legendre polynomials. The volume-preserving condition eliminates the $P_0$ mode, and at leading order the surface is $r_s = a[1 + \eps c_2 P_2(\cos\theta)]$ with
\begin{equation}
\eps\,c_2 = \frac{2}{3}\delta + \Order{\delta^2} = \frac{2}{3}\!\left(\frac{c}{b} - 1\right) + \Order{\left(\frac{c}{b}-1\right)^2}\!.
\label{eq:spheroid_map}
\end{equation}
The eccentricity $e = \sqrt{1 - b^2/c^2}$ is related to $\delta$ by $e^2 = 2\delta + \Order{\delta^2}$, so
\begin{equation}
\eps\,c_2 = \frac{e^2}{3} + \Order{e^4}.
\label{eq:epsc2_vs_e}
\end{equation}

We retain only the $P_2$ mode of the spheroid surface in this mapping. As pointed out to us, this is \emph{not} an additional approximation. The Legendre expansion of the spheroid surface (\ref{eq:spheroid_surface}) contains, besides $P_0$, only even harmonics $P_2, P_4, P_6, \ldots$ by its up--down symmetry; the $P_0$ mode is removed by the equal-volume constraint, and every harmonic $P_n$ with $n \geq 4$ is electrophoretically silent at $\Order{\eps}$ (\S\,\ref{sec:silencing}) and so contributes nothing to the mobility regardless of its amplitude. Consequently the only approximations entering the comparison are the truncation of the perturbation series at $\Order{\eps}$ and the linearisation $\eps c_2 = \tfrac{2}{3}\delta + \Order{\delta^2}$ of the geometric mapping; at this order the $P_2$ result is exact, not merely a small-$\delta$ approximation.

\subsubsection{H\"{u}ckel-limit check}

In the H\"{u}ckel limit ($\kap a \to 0$), \citet{yoon1989} obtained exact expressions for the parallel and perpendicular mobility coefficients $C_\parallel$ and $C_\perp$ of a prolate spheroid of arbitrary eccentricity (their equations~42 and~43). A remarkable property of their result is that the orientation-averaged mobility $C^{\rm AV} = (C_\parallel + 2C_\perp)/3 = 2/3$ \textit{exactly} for all eccentricities (their equation~44). That is, the H\"{u}ckel-limit mobility of a randomly oriented spheroid is the same as that of a sphere, regardless of aspect ratio.

The exact parallel mobility in the H\"{u}ckel limit is (Yoon \& Kim, equation~42)
\begin{equation}
C_\parallel = \frac{-e + (1 + e^2)\,\mathrm{atanh}(e)}{2e^2\,\mathrm{atanh}(e)}.
\label{eq:yk_Cpar_exact}
\end{equation}
We expand for small $e$ using $\mathrm{atanh}(e) = e + e^3/3 + e^5/5 + \Order{e^7}$. The numerator is
\[
-e + (1+e^2)(e + e^3/3 + \cdots) = \tfrac{4}{3}e^3 + \tfrac{8}{15}e^5 + \Order{e^7},
\]
and the denominator is $2e^2(e + e^3/3 + \cdots) = 2e^3(1 + e^2/3 + \cdots)$. Dividing,
\begin{equation}
C_\parallel = \frac{2}{3}\!\left(1 + \frac{e^2}{15} + \Order{e^4}\right).
\label{eq:yk_huckel_expansion}
\end{equation}
Our perturbation result, combined with the mapping (\ref{eq:epsc2_vs_e}), predicts
\[
C_\parallel = \frac{2}{3}\bigl(1 + \eps\,c_2\,\sigma_2(0)\bigr) = \frac{2}{3}\!\left(1 + \frac{e^2}{3}\cdot\frac{1}{5}\right) = \frac{2}{3}\!\left(1 + \frac{e^2}{15}\right),
\]
in exact agreement with (\ref{eq:yk_huckel_expansion}). This consistency check simultaneously validates the Brenner drag coefficient $\alpha_2^{\rm drag} = 1/5$ and the spheroid-to-$P_2$ mapping. Note that the same expression applies to oblate spheroids ($c < b$) through the mapping (\ref{eq:spheroid_map}): since $c/b - 1 < 0$ for an oblate particle, $\eps\,c_2 < 0$ and the correction $\sigma_2(0)\,\eps\,c_2 < 0$, yielding $C_\parallel < 2/3$. The perturbation formula thus naturally recovers the physically expected result that a prolate spheroid aligned with the field moves faster than a sphere, while an oblate one moves slower.

\subsubsection{Full $\kap a$ comparison}

\begin{figure}
\centering
\includegraphics[width=\textwidth]{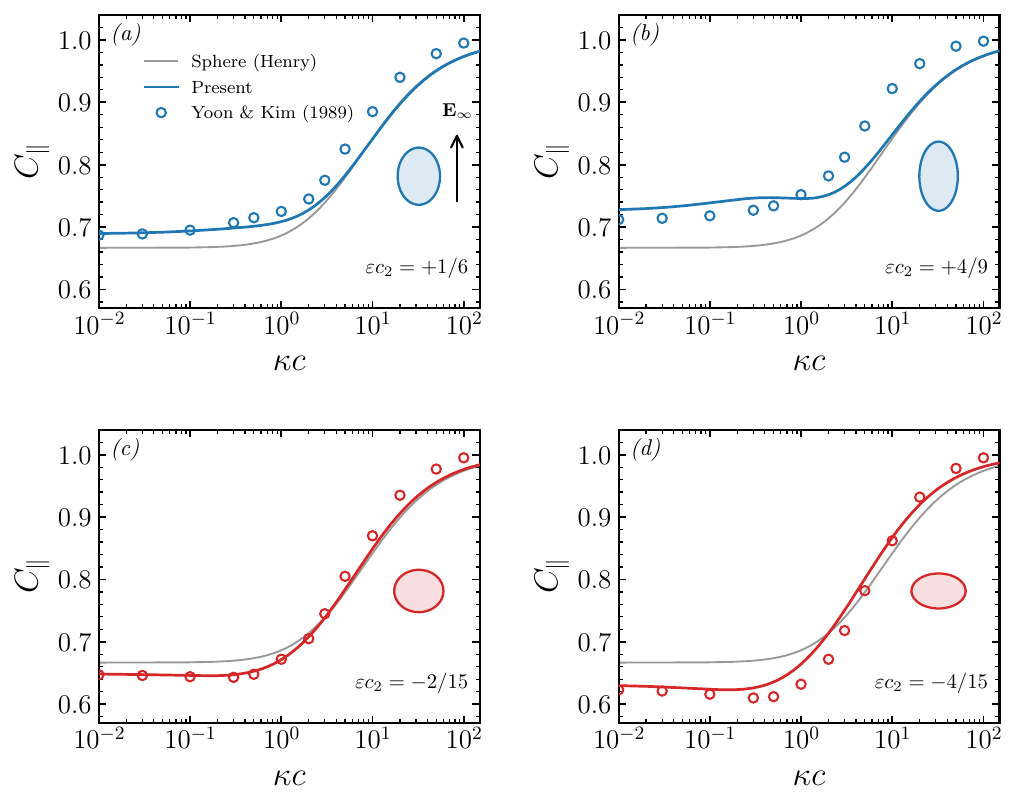}
\caption{Comparison of the perturbation theory (solid lines) with the exact spheroid solutions of \citet{yoon1989} (open circles, digitised from their figure~4). Each panel shows the mobility coefficient $C_\parallel(\kap c)$ for a single spheroid geometry (silhouette inset), with the grey line indicating Henry's function for a sphere. ($a$)~Prolate, $c/a = 0.8$, $\eps c_2 = +1/6$. ($b$)~Prolate, $c/a = 0.6$, $\eps c_2 = +4/9$. ($c$)~Oblate, $c/a = 0.8$, $\eps c_2 = -2/15$. ($d$)~Oblate, $c/a = 0.6$, $\eps c_2 = -4/15$. The electric field is along the symmetry axis in all cases. For $c/a = 0.8$ the perturbation theory is nearly indistinguishable from the exact solution; for $c/a = 0.6$ modest $\Order{\eps^2}$ discrepancies appear at intermediate $\kap c$.}
\label{fig:yk}
\end{figure}

To validate the perturbation theory beyond the H\"{u}ckel limit, we compare the predicted parallel mobility with the exact spheroid solutions of \citet{yoon1989} for two aspect ratios: $c/a = 0.8$ (near-spherical) and $c/a = 0.6$ (moderate deformation). The four cases span both prolate and oblate geometries with $\eps c_2$ ranging from $-4/15$ to $+4/9$. Figure~\ref{fig:yk} presents each case in a separate panel, with the spheroid silhouette inset. The \cite{yoon1989} solutions used for comparison are a digitised data set, read from their figure~4.

For $c/a = 0.8$ ($\eps c_2 = +1/6$ prolate, $-2/15$ oblate), the perturbation theory agrees closely with the exact solutions across the entire range of $\kap c$. The prolate mobility increases monotonically from the H\"{u}ckel value $C_\parallel(0) = 0.687$ towards the Smoluchowski limit $C_\parallel = 1$, while the oblate mobility starts below the sphere value ($C_\parallel(0) = 0.647$), exhibits a subtle minimum near $\kap c \sim 0.3$ (consistent with the non-monotonic behaviour noted by \citealt{yoon1989}), and then rises to unity. The minimum arises because an oblate particle with the field along its short axis has both higher drag and a modified electrostatic driving force that initially compounds the drag penalty before the thin-double-layer enhancement takes over. This is the same physics as the overshoot of $\sigma_2$ in figure~\ref{fig:sigma2}, viewed through the sign of the deformation. For an oblate particle $\eps c_2 < 0$, so the positive shape correction $\sigma_2(\kap a)$ enters $C_\parallel = f_H(\kap a)[1 + \eps c_2 \sigma_2(\kap a)]$ as a \emph{reduction} of the mobility, and this reduction is deepest near $\kap a \approx 0.5$ where $\sigma_2$ peaks. At that point $f_H$ is still well below its Smoluchowski value, so the deepening offset wins and $C_\parallel$ dips below its starting value; as $\kap a$ grows, $\sigma_2 \to 0$ and the rising $f_H$ pulls $C_\parallel$ back up towards unity. The shallow minimum is thus the product of a rising $f_H$ and a non-monotonic correction factor, both of which are encoded in figure~\ref{fig:sigma2}.

For $c/a = 0.6$ ($\eps c_2 = +4/9$ prolate, $-4/15$ oblate), the perturbation parameter is no longer small, and modest discrepancies emerge at intermediate $\kap c$, as expected from the neglected $\Order{\eps^2}$ terms. Nonetheless, the perturbation theory correctly captures the qualitative trends: the prolate curve lies above the sphere line, the oblate curve below, and both approach unity at large $\kap c$. The H\"{u}ckel-limit predictions at $c/a = 0.6$ remain within $2\%$ of the exact values ($C_\parallel^{\rm pro} = 0.712$ exact vs.\ $0.726$ predicted; $C_\parallel^{\rm obl} = 0.623$ exact vs.\ $0.631$ predicted), indicating that the linear mapping (\ref{eq:spheroid_map}) introduces errors that grow as $(c/b - 1)^2$.

\subsection{Electrophoretic silencing of higher harmonics}
\label{sec:silencing}

A striking consequence of the perturbation structure is that only the $P_2$ component of a nearly spherical shape affects the electrophoretic mobility at $\Order{\eps}$. Higher harmonics ($P_3$, $P_4$, and beyond) as well as non-axisymmetric modes ($m \neq 0$) are \textit{electrophoretically silent} at leading order: as established term by term in \S\,\ref{sec:higher_modes}, $\sigma_n \equiv 0$ exactly for every $n \geq 3$, not merely as a small correction.

This follows from the angular selection rules governing the volume integrals in (\ref{eq:sigma2_assembly}). The applied field enters through $\nabla\Phi_0$, which has $P_1$ (dipolar) angular symmetry, and the mobility is extracted by projecting the resulting force in the field direction. The coupling between a shape mode $P_n$ and the applied field $P_1$ produces a net $P_1$ component in the body force only when $n = 0$ or $n = 2$. Since $n = 0$ is eliminated by the equal-volume constraint, only the $n = 2$ mode survives. Physically, the $P_1$ electric field cannot ``sense'' a $P_3$ undulation or a $P_4$ waist pinch, because the body-force contributions from opposite lobes cancel upon angular integration.

\begin{figure}
\centering
\includegraphics[width=\textwidth]{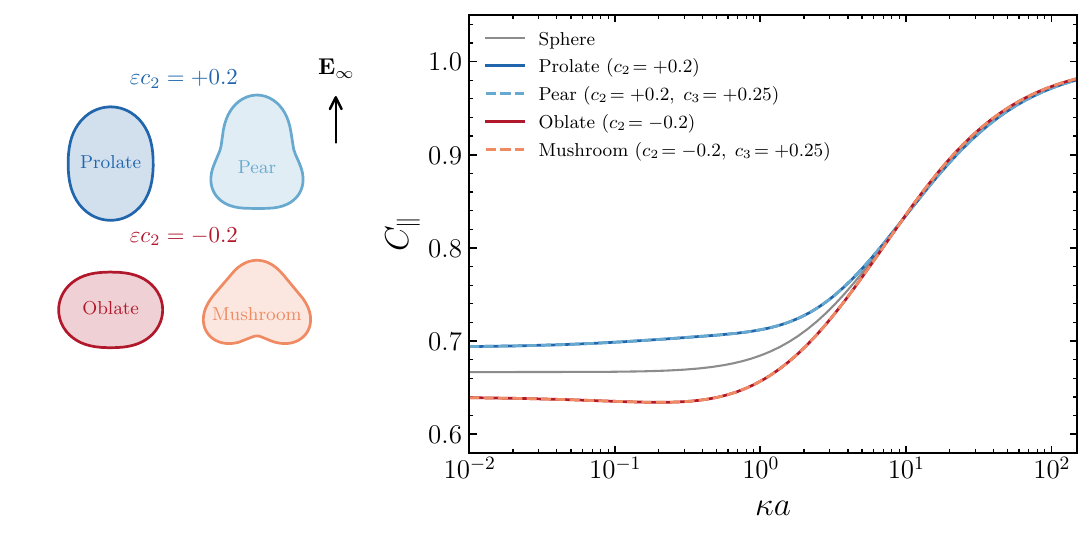}
\caption{Electrophoretic silencing of non-$P_2$ shape modes. \textit{Left:} four nearly spherical particles grouped into two pairs sharing the same $\eps c_2$ but differing in higher harmonics. The prolate ($c_2 = +0.2$) and pear ($c_2 = +0.2$, $c_3 = +0.25$) shapes have the same $P_2$ content; likewise for the oblate ($c_2 = -0.2$) and mushroom ($c_2 = -0.2$, $c_3 = +0.25$) shapes. \textit{Right:} mobility coefficient $C_\parallel(\kap a)$ for each shape. Within each pair, the solid and dashed curves are indistinguishable, confirming that $P_3$ (and higher) modes do not affect the mobility at $\Order{\eps}$.}
\label{fig:silencing}
\end{figure}

To illustrate this prediction, figure~\ref{fig:silencing} compares the mobility of four nearly spherical particles that look markedly different but share the same $P_2$ content. A smooth prolate deformation ($\eps c_2 = +0.2$, $c_n = 0$ for $n \neq 2$) and a pear-shaped particle ($\eps c_2 = +0.2$, $\eps c_3 = +0.25$) have identical $\eps c_2$ and therefore produce indistinguishable mobility curves across the full range of $\kap a$. The same is true for a smooth oblate deformation ($\eps c_2 = -0.2$) and a mushroom-shaped particle ($\eps c_2 = -0.2$, $\eps c_3 = +0.25$), which differ dramatically in appearance (one is top--bottom symmetric with a flattened profile, the other has a protruding cap and a flat base) yet migrate at precisely the same speed.

The two pairs are separated by $|\Delta C/C_{\rm sphere}| \approx 4\%$ in the H\"{u}ckel limit, rising to $\approx 5\%$ near $\kap a \approx 0.5$ where $\sigma_2$ peaks, and collapsing onto the sphere result for $\kap a \gtrsim 50$, consistent with $\sigma_2 \to 0$ in the Smoluchowski limit.

This electrophoretic silencing has a practical implication: the mobility of a nearly spherical particle is insensitive to fine-grained surface roughness, localised protrusions, or top--bottom asymmetry, provided these features project weakly onto $P_2$. What matters for the leading-order shape correction is only the quadrupolar elongation or flattening of the particle. Higher-order shape complexity is filtered out by the angular structure of the electrokinetic problem and can contribute only through $\Order{\eps^2}$ mode-coupling terms.

\begin{figure}
\centering
\includegraphics[width=\textwidth]{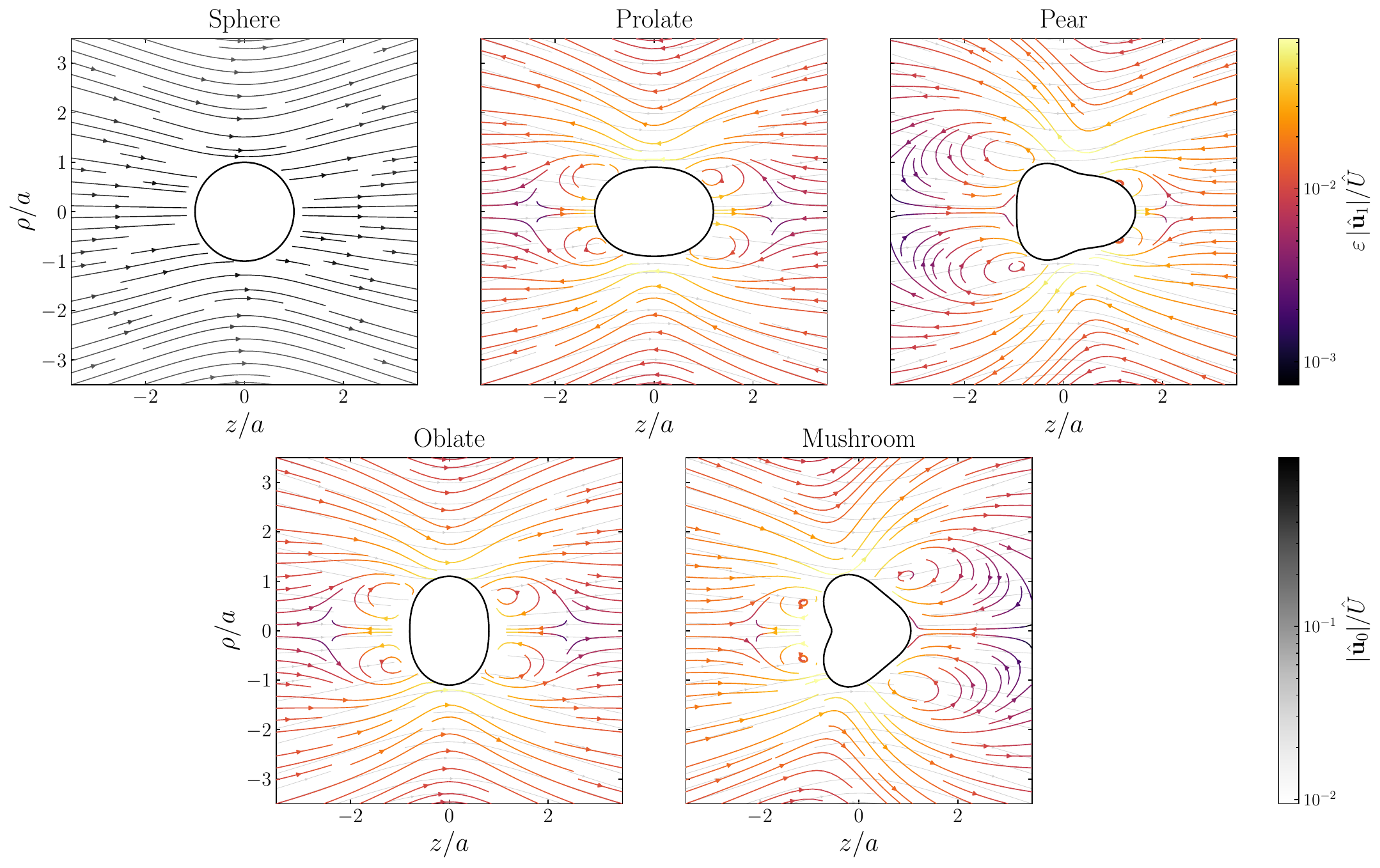}
\caption{Stokes disturbance velocity fields for a sphere and four nearly spherical particles, plotted in the meridional plane with the symmetry axis $z/a$ (direction of the applied field) along the horizontal and the radial coordinate $\rho/a$ along the vertical. In each non-sphere panel, two layers are overlaid: the leading-order base flow $\hat{\mathbf{u}}_0$ (light grey streamlines, common to all shapes) and the $\Order{\eps}$ perturbation $\varepsilon\,\hat{\mathbf{u}}_1$ (coloured streamlines, \textit{inferno} colourmap, logarithmic speed scale). The sphere panel shows $\hat{\mathbf{u}}_0$ alone (greyscale). The prolate and pear shapes share $\eps c_2 = +0.2$ and therefore have the same electrophoretic mobility; likewise the oblate and mushroom share $\eps c_2 = -0.2$. Despite identical mobilities within each pair, the perturbation streamline patterns differ visibly due to the $P_3$ content of the pear and mushroom ($c_3 = +0.25$). These velocity fields are purely hydrodynamic and independent of $\kap a$.}
\label{fig:disturbance_flows}
\end{figure}

Although the electrophoretic mobility is blind to non-$P_2$ shape content, the underlying Stokes disturbance flow is not. Figure~\ref{fig:disturbance_flows} shows the total disturbance velocity $\hat{\mathbf{u}} = \hat{\mathbf{u}}_0 + \eps\,\hat{\mathbf{u}}_1$ for the same four shapes from figure~\ref{fig:silencing}, together with a reference sphere. Within each same-mobility pair (prolate/pear and oblate/mushroom), the streamlines differ noticeably: the $P_3$ content of the pear and mushroom shapes breaks the fore--aft symmetry of the flow, producing asymmetric recirculation patterns that are absent in their pure-$P_2$ counterparts. These differences would be observable in, for instance, tracer-particle trajectories around the migrating colloid, even though the colloid velocity itself is identical. A more detailed visualisation of the individual perturbation fields ($\eps\psi_1$, $\eps\Phi_1$, and $\eps\hat{\mathbf{u}}_1$) for representative shapes is provided in Appendix~\ref{sec:perturbation_fields}.

\section{Conclusions}
\label{sec:conclusions}

We have derived the electrophoretic mobility of a nearly spherical particle at arbitrary Debye length, restricted to weakly charged particles (small zeta potential, $e\zeta/k_BT \ll 1$) and weak applied fields, which are the assumptions that underpin the linearised Poisson--Boltzmann framework and the decomposition of the problem into independent sub-problems used throughout, expressing the result in the compact form $C_\parallel = f_H(\kap a)\,[1 + \eps\,c_2\,\sigma_2(\kap a)]$, where $f_H$ is Henry's function for a sphere and $\sigma_2(\kap a)$ is a universal shape correction coefficient that encodes the entire effect of $P_2$ deformation on particle migration. The principal findings are as follows.

\begin{enumerate}
\item The shape correction $\sigma_2(\kap a)$ interpolates smoothly between the H\"{u}ckel value $\sigma_2(0) = +1/5$, set entirely by the Brenner drag correction, and the Smoluchowski limit $\sigma_2(\infty) = 0$, where the Morrison--Teubner theorem is recovered self-consistently through a cancellation of the $\Phi_1$ and drag contributions.

\item The perturbation theory agrees quantitatively with the exact spheroid solutions of \citet{yoon1989} for $c/a = 0.8$, in both the prolate and oblate orientations, across the full range of $\kap a$. In particular, the subtle non-monotonic behaviour of the oblate parallel mobility at intermediate $\kap a$ is captured.

\item At $\Order{\eps}$, only the $P_2$ component of the particle shape contributes to the electrophoretic mobility. Higher Legendre modes ($P_3$, $P_4$, \ldots) and non-axisymmetric deformations are electrophoretically silent, as demonstrated by comparing pairs of particles that look markedly different (prolate versus pear-shaped, oblate versus mushroom-shaped) but produce indistinguishable mobility curves when they share the same $P_2$ content. This is a consequence of angular selection rules governing the coupling between the dipolar applied field and the shape perturbation.
\end{enumerate}

While we have focused on electrophoresis, the theoretical framework based on the unified mobility expression of \citet{ganguly2024} is readily applicable to other phoretic phenomena. For electrolytic diffusiophoresis, the equilibrium potential $\psi_0$ and applied field $\Phi_0$ are modified by the diffusion-potential contribution arising from unequal ionic diffusivities \citep{prieve1984, gupta2019}, but the perturbation structure in shape remains identical: the same $\sigma_2(\kap a)$ decomposition applies, with only the base-state fields replaced. For non-electrolytic diffusiophoresis, where the body force arises from solute--surface interactions rather than ionic screening, the analysis extends analogously through the appropriate thin-layer or volume-integral formulation \citep{anderson1989}.

A further direction of considerable interest is the connection to self-propelling particles. The rapidly growing field of active colloids, in which particles generate local chemical or thermal gradients to achieve autonomous motion \citep{paxton2004, golestanian2007}, introduces additional complexity through asymmetric surface activity. Prior work from our group has analysed the self-phoretic propulsion of bent-rod particles using slender-body theory \citep{ganguly2023bentrod, raj2023, ganguly2023diffusiophoresis}, where both the geometry and the catalytic patch break the fore--aft symmetry that governs propulsion speed and direction. The present perturbation approach for shape effects could be combined with the active-patch formulation to investigate how particle shape and activity distribution jointly control self-propulsion at finite Debye length. More broadly, the framework utilized in this manuscript is well suited to probe finite interaction effects, for instance particle--wall or particle--particle hydrodynamic and electrokinetic coupling, while simultaneously accounting for non-spherical shape and non-uniform surface activity.

\section{Usage of AI assistance}
\label{sec:ai}

\begin{figure}
  \centerline{\includegraphics[width=\textwidth]{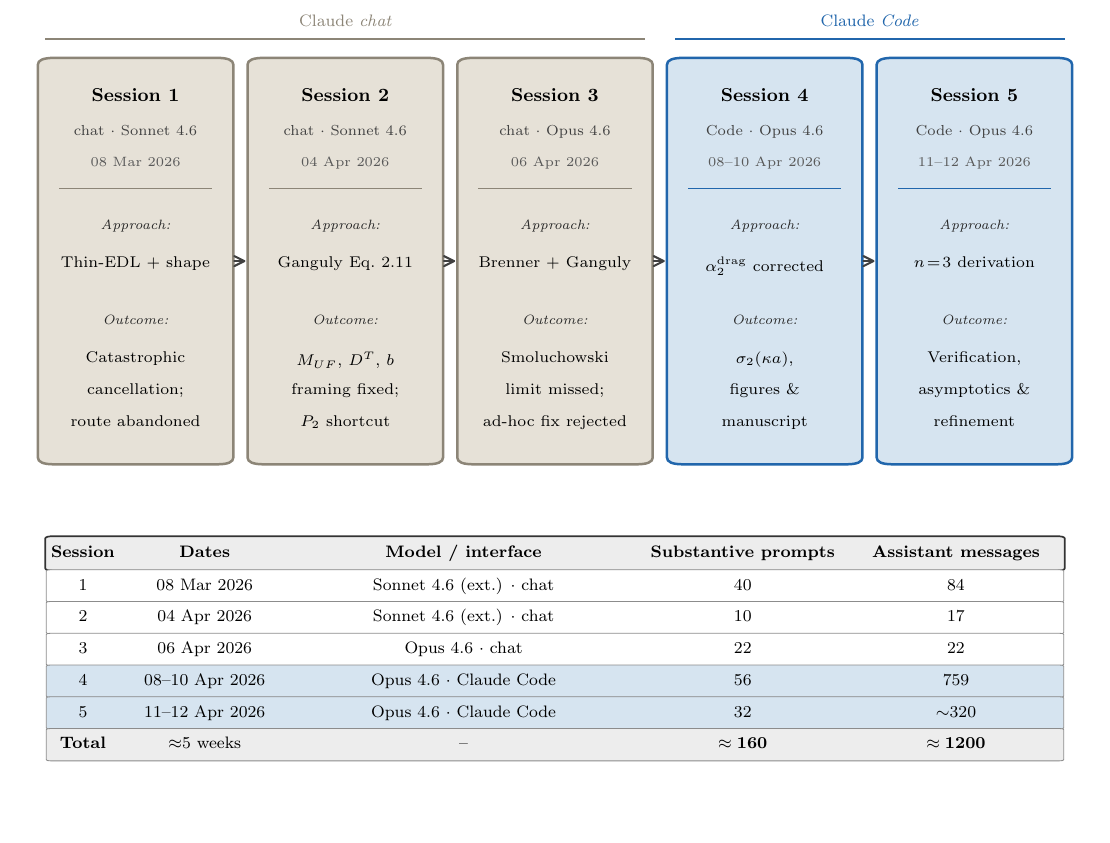}}
  \caption{Summary of the AI-assisted workflow used in developing this manuscript. Three successive sessions in the Claude \textit{chat} interface (two with Sonnet~4.6 extended, one with Opus~4.6) each explored a distinct mathematical route, with the approaches and outcomes indicated; the work was then transitioned to Claude \textit{Code} (Opus~4.6) over two sessions for computation, verification, figure preparation, and manuscript development. The lower table lists the dates, model/interface, and approximate counts of substantive user prompts and assistant messages for each phase. Note that the large assistant-message counts in Sessions~4 and~5 reflect the fact that Claude Code emits one message per tool call (file read, edit, command execution), so the counts overstate the amount of genuine dialogue relative to the chat sessions.}
  \label{fig:ai_workflow}
\end{figure}

This manuscript was developed with the assistance of Claude (Anthropic) over a period of approximately one month, through three successive sessions in the Claude chat interface followed by two development phases in Claude Code integrated with VS Code. Expert verification by the authors was carried out at every stage. The initial idea was conceived approximately 1.5 years ago as a calculation of electrolytic diffusiophoretic mobilities of spheroidal particles, inspired by the work of \citet{yoon1989}, and was later broadened to a general treatment of shape effects. A tractable formulation did not emerge, until a colleague (Prof.~H.~A.~Stone) suggested focusing on the deformed sphere as the canonical shape perturbation. Figure~\ref{fig:ai_workflow} summarises the five-phase workflow; in total, the authors issued approximately $160$ substantive prompts across the five phases.

The three chat sessions (two with Sonnet~4.6 extended, and one with Opus~4.6) each explored a different mathematical route. The first session attempted a double asymptotic expansion in the thin-EDL (Stone--Samuel) limit combined with a small-shape perturbation. As documented in Appendix~\ref{sec:prompts}, this route revealed two structural difficulties. First, two individually $O(\kap a)$ contributions were expected to cancel to leave an $O(1/\kap a)$ residual, which is catastrophic under floating-point subtraction. Second, the model repeatedly constructed post-hoc justifications for intermediate results rather than questioning their validity. Pressing the model to explain a factor of $-9/5$ that it had produced eventually led to the acknowledgment that the calculation had not converged. Only after the authors proposed relaxing the thin-EDL assumption entirely, in favour of a Debye--H\"{u}ckel linearisation with a single perturbation in $\varepsilon$, did a viable direction emerge.

The second session was framed directly around the \citet{ganguly2024} reciprocal-theorem formulation, with the four components $M_{UF}$, $D^T$, $b$, and $\mathrm{d}V$ identified as the quantities requiring perturbation to $O(\varepsilon)$. An early misconception of one of the authors, that the self-mobility $M_{UF}$ remains unchanged to $O(\varepsilon)$, was caught after examining \citet{brenner1964}, which provides the required shape correction to the Stokes drag. The third session consolidated this insight, using Brenner's result together with the Ganguly et al.\ framework to set up the full calculation. Even at this stage, however, the model continued to struggle. The Smoluchowski limit was not recovered in the large-$\kap a$ regime, and an interim ``fix'' involving an ad-hoc factor of $0.8$ was proposed before the authors flagged the interpolation as unacceptable and insisted on a full calculation. It was at this point that the work was transitioned into Claude Code, to gain finer control over file management and iterative manuscript development.

The first Claude Code phase, using Opus~4.6, was the most productive. During this phase, a sign/magnitude error in the Brenner drag coefficient ($\alpha_2^{\rm drag} = +1/5$, not $-3/5$) was identified and corrected, the correct shape correction coefficient $\sigma_2(\kap a)$ was computed and validated against the exact Y\&K solutions, and all figures and manuscript text were developed. A second Claude Code phase, also using Opus~4.6, focused on refinement and verification. The $n = 3$ Gegenbauer streamfunction decomposition was derived and symbolically verified, the sign convention in the streamfunction definition (Convention~A of Kim \& Karrila) was audited and corrected, the perturbation field visualisation (Figure~\ref{fig:perturbation_fields}) was produced, and the asymptotic analysis of the Smoluchowski and H\"uckel limits was expanded.

Throughout the project, the authors performed a systematic series of consistency checks at each stage. Whenever possible, key results were verified independently by hand: the Brenner drag coefficient $\alpha_2^{\rm drag} = +1/5$ was re-derived from equation~1.7 of \citet{brenner1964}, the H\"uckel-limit expansion was matched term-by-term against equation~42 of \citet{yoon1989} (recovering the $e^2/15$ coefficient), and the sign convention in the Gegenbauer streamfunction definition was audited against \citet{kim2005}. Beyond these by-hand checks, the authors read and audited the AI-generated Python code, ran symbolic verification scripts (using SymPy) to confirm governing equations, boundary conditions, and the biharmonic property $\mathrm{E}^4\Psi_1 = 0$, and performed physical-intuition checks (e.g.\ confirming that prolate particles move faster and oblate ones slower), in addition to the cross-validation against \citet{yoon1989} and the limiting cases discussed above. The raw Python scripts that perform the symbolic derivations and numerical computations are provided in the source distribution for full transparency. Representative prompts from each of the five phases are provided in Appendix~\ref{sec:prompts}.

The authors see substantial potential in AI-assisted theoretical research, but also clear limitations. Tasks that are time-intensive but well-specified (assembling perturbation integrals, implementing boundary-value solvers, drafting manuscript prose, and producing publication-quality figures) benefited greatly from AI assistance. At the same time, the chain of reasoning produced by the model had to be parsed carefully to ensure self-consistency, and the model tended to construct justifications for its own answers that had to be actively challenged. Physical reasoning, problem formulation, identification of the correct asymptotic regime, and interpretation of results required domain expertise that the AI could not reliably supply on its own. Because the model can produce mistakes that are superficially convincing, these tools must be used thoughtfully and not as a substitute for physical intuition, a point of particular importance in graduate research training. The ability to formulate the right question, recognise an inconsistency, and locate the source of an error remains squarely within the domain of the researcher.

\section*{Supplementary material}

The following files, generated by Claude during the development of this manuscript, are provided as supplementary material. These files are shared exactly as produced by the AI, without any manual modifications by the authors, so that the reader can independently assess the correctness and utility of AI-generated code.

\begin{itemize}
\item \texttt{fig1\_schematic.tex} --- Standalone Ti\textit{k}Z source for the problem schematic (Figure~1), showing the deformed particle, coordinate system, Debye layer, and applied field.

\item \texttt{plot\_results.py} --- Computes $\sigma_2(\kap a)$ and its decomposition (Figure~2), generates the comparison with \citet{yoon1989} (Figure~3), and produces the electrophoretic silencing figure (Figure~4).

\item \texttt{derive\_u1\_n3.py} --- Symbolic derivation and verification (using SymPy) of the $n = 3$ Gegenbauer streamfunction decomposition, including the radial functions $h_3(r)$ and $h_5(r)$, the velocity components, and the perturbed applied potential $\Phi_1^{(n=3)}$.

\item \texttt{verify\_perturbation\_fields.py} --- Symbolic cross-checks of the $n = 2$ perturbation fields (equations~3.7, 3.14, 3.31--3.32) verifies the governing PDEs, boundary conditions, incompressibility, and the biharmonic streamfunction property $\mathrm{E}^4\Psi_1 = 0$.

\item \texttt{fig\_perturbation\_fields.py} --- Generates the $4 \times 3$ visualisation of the $\Order{\eps}$ residual perturbation fields $\psi_1$, $\Phi_1$, and $\hat{\mathbf{u}}_1$ for four representative shapes (Figure~7).

\item \texttt{fig\_disturbance\_flows.py} --- Generates the Stokes disturbance velocity figure with overlaid base flow and perturbation streamlines for five shapes (Figure~5).

\item \texttt{fig\_ai\_workflow.py} --- Generates the AI-assisted workflow summary figure with session statistics (Figure~6).
\end{itemize}

\section*{Acknowledgements}

A.G.\ thanks the U.S.\ National Science Foundation (CBET-2238412, CAREER award) and the Air Force Office of Scientific Research (FA9550-25-1-0176, Young Investigator Program award) for financial support. The authors are grateful to Professor Howard A.\ Stone for suggesting the deformed-sphere approach during a dinner conversation with A.G., which catalysed the formulation of the present problem. The members of the LIFE research group are thanked for reading the manuscript and providing feedback.

\section*{Declaration of interests}

The authors report no conflict of interest.

\bibliographystyle{jfm}
\bibliography{jfm}

\appendix
\section{Visualisation of the $\Order{\eps}$ perturbation fields}
\label{sec:perturbation_fields}

The analytical expressions for the $\Order{\eps}$ perturbation fields derived in \S\,\ref{sec:psi1}--\ref{sec:u1} can be visualised directly to build physical intuition for how the deformed-sphere fields adapt to the imposed shape. Although the mobility correction $\sigma_n(\kap a)$ vanishes exactly for every $n \geq 3$ (\S\,\ref{sec:silencing}), the underlying \emph{fields} $\psi_1$, $\Phi_1$ and $\hat{\mathbf{u}}_1$ inherit the full angular content of $f(\theta)$ and are therefore non-trivial for any choice of $\{c_n\}$. To make this point concretely we evaluate the residual fields $\eps\psi_1$, $\eps\Phi_1$ and $\eps\hat{\mathbf{u}}_1$ for four representative shape modes,
\begin{equation}
f(\theta) = c_2 P_2(\cos\theta) + c_3 P_3(\cos\theta),
\label{eq:appA_shape}
\end{equation}
namely a prolate ($\eps c_2 = +0.15$, $\eps c_3 = 0$), an oblate ($\eps c_2 = -0.15$, $\eps c_3 = 0$), a pear ($\eps c_2 = +0.10$, $\eps c_3 = +0.10$) and a mushroom ($\eps c_2 = -0.10$, $\eps c_3 = +0.15$). Each field is linear in the $c_n$, so the $n = 2$ and $n = 3$ contributions simply superpose; the $n = 2$ closed forms are (\ref{eq:psi1_sol}), (\ref{eq:Phi1_sol}) and (\ref{eq:u1r})--(\ref{eq:u1th}), while the $n = 3$ closed forms (which require a separate Gegenbauer decomposition) are derived in Appendix~\ref{sec:n3_derivation} below.

Figure~\ref{fig:perturbation_fields} shows the result, with rows indexing the four shapes and columns the three perturbation fields; the white disc in each panel is the reference sphere $r = a$ and the solid curve is the actual deformed surface $r = a[1 + \eps f(\theta)]$. Several features are worth highlighting. First, $\psi_1$ (left column) is sharply confined to a Debye layer of thickness $\sim (\kap a)^{-1}$ and is essentially controlled by the local shape distortion. The prolate and oblate particles are mirror images, while pear and mushroom break the up--down symmetry. This confinement is the geometric origin of the $\eps\kap a$ ordering of the perturbation expansion. Second, $\Phi_1$ (middle column) decays only algebraically; its lobes also reflect the shape, but the angular structure mixes $(P_1, P_3)$ for the $n = 2$ contribution and $(P_2, P_4)$ for the $n = 3$ contribution. By the angular selection rule of \S\,\ref{sec:silencing}, only the $P_1$ part of $\Phi_1$ couples to the mobility at $\Order{\eps}$; everything else is electrophoretically silent yet clearly visible in the field. Third, $\hat{\mathbf{u}}_1$ (right column) shows the perturbed Stokes disturbance flow as streamlines coloured by the local speed (log scale). For pure $n = 2$ deformations the near-field structure is a $m = 2$ Stokeslet plus $m = 4$ octupole, generating the four-eddy pattern visible in the prolate and oblate panels. For $n = 3$ deformations the corresponding Gegenbauer modes are $m = 3$ and $m = 5$, neither of which carries a Stokeslet (as it should, since a $P_3$ deformation produces no $\Order{\eps}$ drag correction); the resulting streamline pattern is visibly more localised, and is asymmetric under $z \to -z$.

The take-away is that the angular selection rules silence the higher modes only at the level of the volume integral that defines the mobility; the underlying physical fields know about every harmonic of $f(\theta)$, and the shape sensitivity that survives in the mobility is concentrated in the unique $P_2$ component of $f$.

\begin{figure}
\centering
\includegraphics[width=\textwidth]{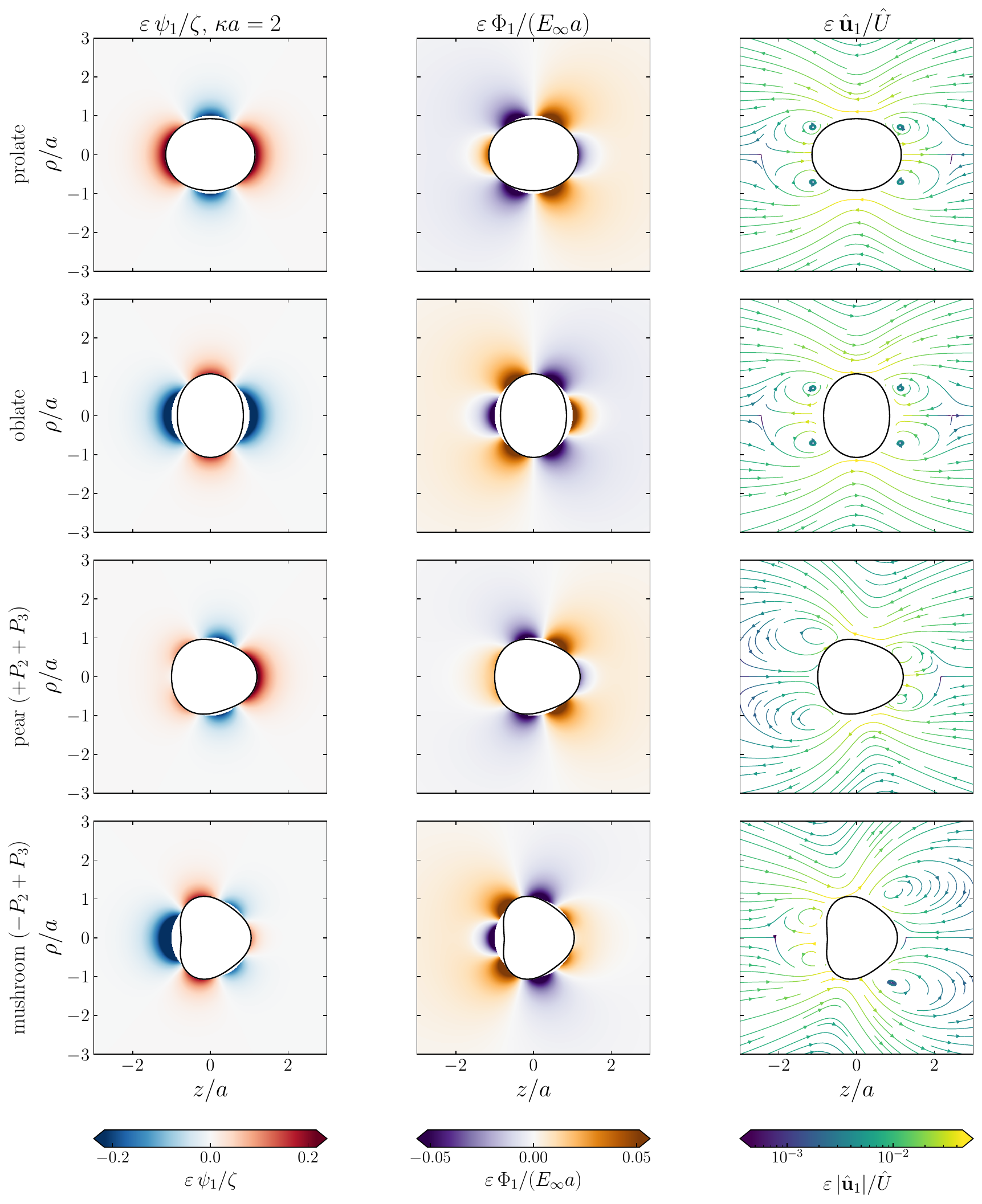}
\caption{Residual $\Order{\eps}$ perturbation fields for four representative shape modes $f(\theta) = c_2 P_2 + c_3 P_3$, plotted in the meridional half-plane with the symmetry axis $z/a$ (direction of the applied field) along the horizontal and the radial coordinate $\rho/a$ along the vertical. Rows: prolate ($\eps c_2 = +0.15$), oblate ($\eps c_2 = -0.15$), pear ($\eps c_2 = +0.10$, $\eps c_3 = +0.10$), mushroom ($\eps c_2 = -0.10$, $\eps c_3 = +0.15$); the solid black curve in every panel is the deformed surface $r = a[1 + \eps f(\theta)]$. Columns: \emph{(left)} the residual perturbed equilibrium potential $\eps\psi_1/\zeta$ from (\ref{eq:psi1_sol}) at $\kap a = 2$; \emph{(middle)} the residual perturbed applied potential $\eps\Phi_1/(E_\infty a)$ from (\ref{eq:Phi1_sol}) and its $n = 3$ analogue (\ref{eq:Phi1_n3_sol}); \emph{(right)} streamlines of the residual Stokes disturbance velocity $\eps\hat{\mathbf{u}}_1/\hat{U}$ from (\ref{eq:u1r})--(\ref{eq:u1th}) and (\ref{eq:u1r_n3})--(\ref{eq:u1th_n3}), coloured by the local speed (logarithmic scale). Each column shares a common colour range so that magnitudes can be compared across rows. The $\Phi_1$ and $\hat{\mathbf{u}}_1$ fields are purely hydrodynamic/electrostatic and independent of $\kap a$; only $\psi_1$ depends on $\kap a$ (shown here at $\kap a = 2$). The fields are linear in the $c_n$, so the contributions from each shape mode superpose.}
\label{fig:perturbation_fields}
\end{figure}

\section{Closed form for $\hat{\mathbf{u}}_1$ in the $n = 3$ mode}
\label{sec:n3_derivation}

The reduction of the perturbation problem for a $P_2$ shape mode to the closed-form expressions (\ref{eq:u1r})--(\ref{eq:u1th}) is carried out in \S\,\ref{sec:u1}. For the visualisation in Appendix~\ref{sec:perturbation_fields} we also need the analogous closed forms for a $P_3$ shape mode, $f(\theta) = c_3 P_3(\cos\theta)$; we record them here together with their derivation. The mobility correction $\sigma_3 \equiv 0$ irrespective of $\kap a$ (\S\,\ref{sec:silencing}), so the result is not needed for the main calculation, but it is needed if one wants to actually plot the perturbation fields for an asymmetric particle.

The boundary condition follows the same Taylor expansion as for the $n = 2$ mode: in the body frame, $\hat{\mathbf{u}}_1$ must vanish on the deformed surface, which translates to $\hat{u}_{1,r}(a,\theta) = 0$ and
\begin{equation}
\hat{u}_{1,\theta}(a,\theta) = -\tfrac{3}{2}\,\hat{U}\,c_3\,P_3(\cos\theta)\,\sin\theta.
\label{eq:u1th_BC_n3}
\end{equation}
We seek a stream function $\Psi_1$ for axisymmetric Stokes flow satisfying $\mathrm{E}^4 \Psi_1 = 0$ in $r > a$ with $\Psi_1(a,\theta) = 0$ (impenetrability) and $\partial_r\Psi_1|_{r=a} = (3/2)\hat{U}c_3 a\sin^2\theta\, P_3(\cos\theta)$, the latter obtained by inverting $\hat{u}_{1,\theta} = -(r\sin\theta)^{-1}\partial_r\Psi_1$. The natural angular basis is the Gegenbauer functions of order $-1/2$,
\begin{equation}
C_m^{-1/2}(\eta) = \frac{P_{m-2}(\eta) - P_m(\eta)}{2m-1}, \qquad \eta = \cos\theta,
\end{equation}
which diagonalise the operator $\mathrm{E}^2$ in the form $\mathrm{E}^2[h_m(r) C_m^{-1/2}] = [h_m'' - m(m-1) h_m/r^2]\,C_m^{-1/2}$. Decomposing the angular forcing $\sin^2\theta\, P_3$ in this basis yields
\begin{equation}
\sin^2\!\theta\,P_3(\cos\theta) = -\tfrac{6}{7}\,C_3^{-1/2}(\cos\theta) + \tfrac{20}{7}\,C_5^{-1/2}(\cos\theta),
\label{eq:Gegenbauer_split}
\end{equation}
so $\Psi_1 = h_3(r) C_3^{-1/2} + h_5(r) C_5^{-1/2}$ with $h_3$ and $h_5$ each satisfying $[\mathrm{d}^2/\mathrm{d}r^2 - m(m-1)/r^2]^2 h_m = 0$. Of the four power-law solutions for each mode, only those that are bounded at infinity and produce decaying velocity fields survive: for $m = 3$ these are $r^0 = 1$ (bounded) and $r^{-2}$ (decaying), and for $m = 5$ they are $r^{-2}$ and $r^{-4}$. Note that no Stokeslet ($r$) or source-dipole ($r^{-1}$) modes appear, consistent with Brenner's result that a $P_3$ shape distortion does not perturb the Stokes drag at $\Order{\eps}$. Imposing $h_m(a) = 0$ and matching $h_m'(a)$ to the Gegenbauer projection of the boundary forcing fixes the constants uniquely:
\begin{equation}
h_3(r) = \frac{9}{14}\,\hat{U}\,c_3\,a^2\,\frac{a^2 - r^2}{r^2}, \qquad
h_5(r) = \frac{15}{7}\,\hat{U}\,c_3\,a^4\,\frac{r^2 - a^2}{r^4}.
\label{eq:h35_n3}
\end{equation}
The resulting stream function is
\begin{equation}
\Psi_1^{(n=3)}(r,\theta) = h_3(r)\,C_3^{-1/2}(\cos\theta) + h_5(r)\,C_5^{-1/2}(\cos\theta),
\label{eq:Psi1_n3}
\end{equation}
and the radial velocity component, obtained via $\hat{u}_{1,r} = (r^2\sin\theta)^{-1}\partial_\theta\Psi_1$ together with the identity $\partial_\theta C_m^{-1/2}(\cos\theta) = \sin\theta\,P_{m-1}(\cos\theta)$, takes the compact Legendre form (in the dimensionless radius $s = r/a$)
\begin{equation}
\hat{u}_{1,r}^{(n=3)} = \hat{U}\,c_3\left[\frac{9}{14}\!\left(\frac{1}{s^4} - \frac{1}{s^2}\right) P_2(\cos\theta) + \frac{15}{7}\!\left(\frac{1}{s^4} - \frac{1}{s^6}\right) P_4(\cos\theta)\right].
\label{eq:u1r_n3}
\end{equation}
The polar component $\hat{u}_{1,\theta}^{(n=3)} = -(r\sin\theta)^{-1}\partial_r\Psi_1^{(n=3)}$ does not collapse to a clean Legendre representation but follows immediately from (\ref{eq:Psi1_n3}) by direct differentiation; we use the explicit polynomial form (verified symbolically and reproduced in the figure script) without writing it out here. For brevity we collectively label this pair of components (\ref{eq:u1r_n3}) and the implicit $\hat{u}_{1,\theta}^{(n=3)}$ as
\begin{equation}
\hat{u}_{1,\theta}^{(n=3)}(r,\theta) = -\frac{1}{r\sin\theta}\,\frac{\partial}{\partial r}\!\left[h_3(r)\,C_3^{-1/2} + h_5(r)\,C_5^{-1/2}\right].
\label{eq:u1th_n3}
\end{equation}
The corresponding perturbed applied potential, derived by exactly the same Legendre projection as in \S\,\ref{sec:Phi1} but with $f' = -(3/2) c_3 (5\cos^2\theta - 1)\sin\theta$, is
\begin{equation}
\Phi_1^{(n=3)}(r,\theta) = c_3\,E_\infty\,a\!\left[\frac{3}{7}\left(\frac{a}{r}\right)^{\!3} P_2(\cos\theta) - \frac{6}{7}\left(\frac{a}{r}\right)^{\!5} P_4(\cos\theta)\right],
\label{eq:Phi1_n3_sol}
\end{equation}
and the perturbed equilibrium potential, by the same screened-Helmholtz argument as in \S\,\ref{sec:psi1}, is $\psi_1^{(n=3)} = \zeta c_3 (1 + \kap a)\,k_3(\kap r)/k_3(\kap a)\,P_3(\cos\theta)$. We have verified all of these expressions symbolically. The velocity field is divergence-free, the stream function satisfies $\mathrm{E}^4\Psi_1 = 0$, and the boundary conditions $\hat{u}_{1,r}(a) = 0$ and (\ref{eq:u1th_BC_n3}) hold exactly. The script that performs this derivation and verification is provided in the source distribution as \texttt{derive\_u1\_n3.py}.

\section{Representative prompts used with Claude}
\label{sec:prompts}

The following are  representative prompts issued by the authors to Claude during the development of this manuscript. They have been lightly edited only for mathematical typesetting and length; their content and intent are unchanged. They are reproduced to illustrate the nature of the human--AI interaction, including the process of identifying and correcting errors. Only technical prompts are included; those related to formatting, figure aesthetics, and manuscript styling have been omitted. The prompts are grouped by phase, corresponding to three successive sessions in the Claude chat interface and two subsequent phases in Claude Code.

\subsection{Session 1: thin-EDL exploration with Sonnet~4.6 (extended)}

The first session framed the problem as a thin-EDL expansion with a small-shape perturbation:

\begin{quote}
\textit{``This paper lays out a general framework to solve for mobilities for any particle. Reality is, most people only solve spheres, which is what this paper also does. I am interested in electrophoresis, which is covered in this paper. I want to see if that can be done in thin EDL limit, but where we take one additional order of correction. So imagine if the thin EDL limit is kappa = a/lambda $\gg$ 1, we want the first two non-zero orders for electrophoretic mobility. The leading order has to satisfy the Smoluchowski limit, and the next order is the shape correction. This is a challenging problem to do broadly, so how about we do it for a nearly spherical shape $r = a + \varepsilon f(\theta)$, where $\varepsilon \ll 1$, and $f(\theta)$ is an arbitrary function, and $a$ is the radius. You can assume small potentials to keep things easy.''}
\end{quote}

After many iterations using the thin-EDL (Stone--Samuel) slip-velocity approach, the model arrived at a factor of $-9/5$ for the $\kappa^{-1}$ correction that appeared to be incorrect. When pressed for a derivation of that factor, the model acknowledged an error but was unable to arrive at a converged result, because two individually $O(\kap a)$ terms were diverging and their supposed cancellation was producing nonsense. After observing that the model had quietly moved on without resolving the discrepancy, we pressed it to confront the issue:

\begin{quote}
\textit{``Couldn't resolve it?''}
\end{quote}

\begin{quote}
\textit{``Seems like you have given up?''}
\end{quote}

\begin{quote}
\textit{``I am not sure if you clarified how Yoon and Kim give the $-9/5$ factor. Could you first clarify that?''}
\end{quote}

The response revealed the structural issue: the $-9/5$ result in the thin-EDL limit arises from two individually $O(\kap a)$ contributions that cancel to leave an $O(1/\kap a)$ residual, which is catastrophic under floating-point subtraction. Having identified the mechanism, we proposed a reformulation that avoids the double asymptotic expansion entirely:

\begin{quote}
\textit{``Maybe we can relax the assumption of small kappa. Instead we just use Debye--H\"{u}ckel and expand in epsilon -- could you maybe try that? This way we can solve for all kappa.''}
\end{quote}

In the subsequent iteration the model reverted to the Stone--Samuel formula, requiring an explicit correction:

\begin{quote}
\textit{``I am confused by your approach. The Stone--Samuel formula is only true for the large $\kap a$ limit, and using it for all $\kap a$ is futile. You just need to redo section~3.2 of the attached manuscript, but for a nearly spherical axisymmetric particle. My understanding is that the mobility has a correction of $O(\varepsilon^2)$, so it will stay the same, and hence both $D^T$ and $b$ need to be expanded in $O(\varepsilon)$, and then volume integration needs to happen from the perturbed shape to infinity. This is required to capture the effect of all $\kap a$. I do not understand why Stone and Samuel are being invoked.''}
\end{quote}

Finally, we found it necessary to release the model from the assumption that a specific known result must be recovered:

\begin{quote}
\textit{``I broadly agree that this looks correct. I do not really care if Yoon and Kim, the H\"{u}ckel limit, or the Morrison limits are satisfied yet, because I disagree that $c_\sigma = -9/5$ has to come out as I am not yet convinced. The Morrison limit is also thin-EDL only so I don't think it holds either. I just need to see a plot of $c_\sigma(\kap a)$, and then I want to see its variation.''}
\end{quote}

\subsection{Session 2: reframing via the Ganguly et al.\ formulation (Sonnet~4.6 extended)}

The second session began with a clean restart, framed directly in terms of the \citet{ganguly2024} reciprocal-theorem expression:

\begin{quote}
\textit{``I want to follow this manuscript closely and use equation~2.11 to solve the translational electrophoretic mobility of a nearly axisymmetric spherical particle. This needs to be done in the Debye--H\"{u}ckel limit and weak electric field for arbitrary Debye length; essentially formulating the same problem as in section 3.2 of the manuscript, but for a nearly axisymmetric particle. The electric field should be applied in the $z$-direction, and the particle would be axisymmetric in the $x$-$y$ plane. The shape of the sphere $r = a(1 + \varepsilon f(\theta))$, where $\varepsilon \ll 1$. There are four key components to solve for this: (i) $M_{UF}$, (ii) $D^T$, (iii) $b$, and (iv) $\mathrm{d}V$. I want to get answers until $O(\varepsilon)$. My understanding is that $M_{UF}$ relaxes to a sphere since first non-zero corrections are $O(\varepsilon^2)$. $D^T$ would have a correction on $O(\varepsilon)$, $b$ would have a correction of $O(\varepsilon)$, and $\mathrm{d}V$ I am not sure since the volume would change so that is something I am not confident yet.''}
\end{quote}

An early misconception was identified and corrected during the same session:

\begin{quote}
\textit{``I think I made a mistake in assuming that the mobility $M_{UF}$ is constant upon deformation up to $O(\varepsilon)$. It doesn't seem that this is the case based on the attached manuscript. Can you read this in detail? And confirm if this was the missing link, and perhaps that was causing the blowup? And perhaps this paper already gives the disturbance tensor?''}
\end{quote}

\begin{quote}
\textit{``Does the solution simplify if the shapes just contain $P_2(\cos\theta)$?''}
\end{quote}

\subsection{Session 3: Brenner + Ganguly framework (Opus~4.6)}

The third session consolidated these insights by combining Brenner's deformed-sphere drag result with the Ganguly et al.\ framework:

\begin{quote}
\textit{``Based on the other conversations, find the electrophoretic mobility of a nearly spherical axisymmetric particle with small zeta potentials and an arbitrary Debye length to particle size ratio. Use the work of Brenner 1964 to find the disturbance and mobility of a nearly spherical axisymmetric particle, and use that information in Ganguly et al.\ 2024 framework to obtain an expression.''}
\end{quote}

Even with the improved formulation, the Smoluchowski limit was not being recovered at large $\kap a$, and the model proposed an interim interpolation that the authors rejected:

\begin{quote}
\textit{``So there is an artificial fix by adding a $0.8$ value?''}
\end{quote}

\begin{quote}
\textit{``Before you do that, I still want to understand the issue in the theory. Why are we not recovering the Smoluchowski limit in the large $\kap a$ limit?''}
\end{quote}

\begin{quote}
\textit{``I think it would help if this is fixed fully. Please do the full calculation instead of an interpolated ad-hoc formula. I want to ensure that the Smoluchowski limit is fully satisfied.''}
\end{quote}

At this point the authors transitioned the work to Claude Code for finer control over file management and iterative manuscript development.

\subsection{Session 4: manuscript development with Claude Code (Opus~4.6)}

In Claude Code, the authors identified and corrected a sign/magnitude error in the Brenner drag coefficient, finalised the computation of $\sigma_2(\kap a)$, and produced all figures and text. Representative prompts include:

\begin{quote}
\textit{``I am confused with your terminology. In Brenner 1964 Eq.~1.7, the alpha $= -1/5$, but it is on the right hand side of the formula, and hence the mobility would make it $+1/5$, making the result consistent with the low eccentricity formula of Yoon and Kim?''}
\end{quote}

\begin{quote}
\textit{``In the results section, create two sets of results. First, match with the full profiles of Yoon and Kim 1989 paper (you have access to their results in the Literature folder) for the results that can be compared faithfully with this calculation. My understanding is that they would only be weak variations of the sphere, so choose the appropriate values and try to match the results for a range of $\kap a$ values. Check the Ganguly et al paper to make sure the plot style, font size, label size etc.\ are carried over. Be as close to that style as possible.''}
\end{quote}

\begin{quote}
\textit{``I do not want you to solve the YK problem yourself; instead I want you to just digitize their plots and use the formula in our manuscript to compare for the cases they present. If the limit is not appropriate, choose the closest one and compare.''}
\end{quote}

\begin{quote}
\textit{``Next, make a new result with several different shapes. I am thinking of 4 different shapes. The idea would be to create a plot of 4 different shaped particles within the plot with different colors, and plots of $C_\parallel$ with $\kap a$ where the line color matches the shape. These 4 shapes could be chosen somewhat strategically. I don't have a sense of the physics yet, so perhaps let us first discuss what canonical shapes would be before we dig into making the plots.''}
\end{quote}

\begin{quote}
\textit{``I realized I made a mistake in this idea of cylinder. Could we do some more interesting shapes like an hourglass which is wider on top and bottom, and then thin in the middle. An opposite shape of hourglass, so which is like wide in the middle and thin at the ends. And then maybe like a tri-spinner like shape? Not sure if these are easy, but just curious. First, let us iterate on the shapes.''}
\end{quote}

\begin{quote}
\textit{``Yeah the 2+2 version looks great. Please go ahead and make the calculations, and make the result as visually appealing and similar to other figures as possible. Do add physical explanations and insights in the associated text.''}
\end{quote}

\subsection{Session 5: refinement and verification with Claude Code (Opus~4.6)}

A second Claude Code session focused on deriving the $n = 3$ closed forms, auditing sign conventions, expanding the asymptotic discussion, and performing additional consistency checks against the literature. Representative prompts include:

\begin{quote}
\textit{``In the equation setup, we don't understand if the sign was reversed going from Eq.~3.24 to Eqs.~3.29--3.30. It doesn't change the final result, but it seemed a sign flip in our calculations.''}
\end{quote}

This led to the identification and correction of a sign convention mismatch (Convention~A vs.\ Convention~B for the Gegenbauer streamfunction), in which two sign errors had cancelled, leaving the final velocities correct but the intermediate expressions inconsistent.

\begin{quote}
\textit{``I think if you can expand it and show that it becomes $1 - 3/(\kappa a)$ that would also show that the result recovers Smoluchowski but also the first correction for finite double layer thickness. Do not over-expand. Just a little bit.''}
\end{quote}

\begin{quote}
\textit{``Could you explain the Smoluchowski limit rationale below Eq.~3.3? I did not follow the explanation.''}
\end{quote}

The following prompt illustrates the use of physical-intuition checks to probe the perturbation result:

\begin{quote}
\textit{``Would you say that if $b > c$, then eccentricity squared would become negative and hence we will get a lower mobility? Therefore, the expression is basically valid? Perhaps clarifying that below would be useful so that a reader understands that the consistency of higher mobility for prolate and lower for oblate is satisfied here.''}
\end{quote}

Finally, the asymptotic scaling of the $\psi_1$ and $\hat{\mathbf{u}}_1$ integrals at large $\kap a$ was expanded in the manuscript after the following exchange:

\begin{quote}
\textit{``I am a little confused about how you arrived at $\kappa a/5$ and $-\kappa a/5$ for $\psi_1$ and $\hat{u}_1$ integrals. Could you walk through this calculation a bit more clearly?''}
\end{quote}

\begin{quote}
\textit{``Could you expand the manuscript to highlight these details? Otherwise it can be too hard for a reader to understand.''}
\end{quote}

\end{document}